\documentclass[reprint,aps,prd,amsmath,amssymb,nofootinbib,nobibnotes,longbibliography,floatfix,nobalancelastpage]{revtex4-2}
\usepackage[T1]{fontenc}
\usepackage{lmodern}
\usepackage{mathtools}
\usepackage{graphicx}
\usepackage{booktabs}
\usepackage{multirow}
\usepackage{siunitx}
\usepackage{xspace}
\usepackage[caption=false]{subfig}
\usepackage[dvipsnames,table]{xcolor}
\usepackage{hyperref}
\definecolor{linkcolor}{HTML}{0b5394}
\hypersetup{colorlinks=true,citecolor=linkcolor,linkcolor=linkcolor,urlcolor=linkcolor}
\usepackage[capitalise]{cleveref}
\crefname{figure}{Fig.}{Figs.}\Crefname{figure}{Figure}{Figures}
\crefname{table}{Table}{Tables}\Crefname{table}{Table}{Tables}
\crefname{section}{Sec.}{Secs.}\Crefname{section}{Section}{Sections}
\crefname{subsection}{Sec.}{Secs.}\Crefname{subsection}{Section}{Sections}
\crefname{appendix}{Appendix}{Appendices}\Crefname{appendix}{Appendix}{Appendices}
\crefname{subappendix}{Appendix}{Appendices}\Crefname{subappendix}{Appendix}{Appendices}
\crefname{equation}{Eq.}{Eqs.}\Crefname{equation}{Equation}{Equations}

\graphicspath{{figures/}}

\newcommand{\pt}{\ensuremath{p_\mathrm{T}}\xspace}
\newcommand{\ttbar}{\ensuremath{t\bar{t}}\xspace}
\newcommand{\antikt}{anti-\ensuremath{k_\mathrm{T}}\xspace}
\newcommand{\akfour}{AK4\xspace}
\newcommand{\jetclass}{JetClass\xspace}

\newcommand{\nParBackbone}{67.3}
\newcommand{\nParUnion}{144.3}
\newcommand{\nParJets}{107.4}
\newcommand{\nParEvents}{125.1}
\newcommand{\nMaxPartCv}{521}
\newcommand{\nMaxPartJC}{183}
\newcommand{\nParHeadJC}{19.2}
\newcommand{\nParHeadCv}{36.3}
\newcommand{\stepsConstJets}{\num{974848}}
\newcommand{\stepsFinalJets}{\num{1171456}}
\newcommand{\stepsConstUnion}{\num{1376256}}
\newcommand{\stepsFinalUnion}{\num{1654784}}
\newcommand{\stepsConstEvents}{\num{466944}}
\newcommand{\stepsFinalEvents}{\num{565248}}
\newcommand{\passesJets}{12.0}
\newcommand{\passesUnionJC}{13.7}
\newcommand{\passesUnionCv}{14.4}
\newcommand{\passesEvents}{25.4}
\newcommand{\gpuhUnion}{\num{1040}}
\newcommand{\gpuhJets}{\num{485}}
\newcommand{\gpuhEvents}{\num{581}}

\newcommand{\nJCTrain}{100}
\newcommand{\nJCVal}{5.0}
\newcommand{\nJCTotal}{125}
\newcommand{\nJCTest}{20}
\newcommand{\nCvQCDTrain}{9.2}
\newcommand{\nCvQCDVal}{1.2}
\newcommand{\nCvQCDTest}{1.2}
\newcommand{\nCvBBTrain}{2.0}
\newcommand{\nCvBBVal}{0.25}
\newcommand{\nCvBBTest}{0.25}
\newcommand{\nCvHadTrain}{3.8}
\newcommand{\nCvHadVal}{0.48}
\newcommand{\nCvHadTest}{0.48}
\newcommand{\nCvSemiTrain}{3.9}
\newcommand{\nCvSemiVal}{0.48}
\newcommand{\nCvSemiTest}{0.48}
\newcommand{\nCvLepTrain}{3.9}
\newcommand{\nCvLepVal}{0.48}
\newcommand{\nCvLepTest}{0.48}
\newcommand{\nCvTotal}{28.5}
\newcommand{\nTopTrain}{1.2}
\newcommand{\nTopVal}{0.40}
\newcommand{\nTopTest}{0.40}
\newcommand{\nConstJC}{39}
\newcommand{\fracWindowJC}{0.16}

\begin{document}
\captionsetup{justification=justified,singlelinecheck=false}

\title{OmniX: A first jet and event model}

\author{Anna Hallin}
\author{Gregor Kasieczka}
\author{Henning Rose}
\email[Corresponding author: ]{henning.rose@uni-hamburg.de}
\affiliation{Institut f\"ur Experimentalphysik, Universit\"at Hamburg, 22761 Hamburg, Germany}

\begin{abstract}
Foundation models are a new approach to building machine learning algorithms for particle physics. Their central idea is that a single model can be adapted to a variety of downstream tasks, amortizing training costs. So far, most foundation model prototypes have operated on the constituents of a single jet. This limits their usefulness for physics problems at the scale of full events, such as distinguishing different processes, and restricts their performance, as global correlations are invisible to the model. Here, we introduce OmniX, an architecture that operates on both single jets and whole events within a single model. Using the jet-level JetClass and Top-tagging datasets and the event-level Collide-2V dataset, we demonstrate that OmniX trained jointly on single jets and full events matches or exceeds otherwise equivalent jet-only and event-only models in generation and classification.

\end{abstract}

\maketitle

\section{Introduction}
\label{sec:intro}

The foundation model paradigm~\cite{bommasani2022opportunitiesrisksfoundationmodels} moves machine learning away from models built for a single problem and towards models that are pre-trained once on broad data and then adapted to many downstream tasks. Particle physics is well suited to this approach: many analyses operate on closely related datasets and ask similar questions of them, such as particle identification, tagging, or reconstruction in their various forms. A single pre-trained model can therefore potentially reach strong performance on individual problems with little task-specific fine-tuning.

So far, however, foundation models in particle physics~\cite{Hallin:2025ywf} have focused almost exclusively on jets, and in particular on jet tagging~\cite{Qu:2022mxj,Birk:2024knn,Golling:2024abg,Li:2024htp,Mikuni:2024qsr}. This emphasis is well motivated, since jet tagging is a high-impact task central to many analyses, and it has produced a number of key results, such as first hints at transfer~\cite{Birk:2024knn}, understanding of the scaling properties~\cite{ATL-SOFT-PUB-2026-002,Vigl:2026ppx,Amram:2026zzv} or redefining the state of the art for identifying heavy resonances~\cite{Bhimji:2025isp}.

At the same time, a model restricted to jets cannot capture the global structure of the event or the underlying physics process that produced it. Event-level foundation models have received comparatively little attention, with a couple of notable exceptions~\cite{Kishimoto:2023cys,Wildridge:2024yeg,Ho:2024qyf,Kishimoto:2025sQ,Hsu:2026sww,Mokhtar:2026gdb,Wu:2026qlh,Huang:2026jlv}. The data granularity used in these models ranges from high-level physics reconstruction objects~\cite{Kishimoto:2023cys,Wildridge:2024yeg,Ho:2024qyf,Kishimoto:2025sQ,Hsu:2026sww} and particle flow candidates~\cite{Mokhtar:2026gdb}, to reconstructed tracks and calorimeter clusters~\cite{Huang:2026jlv} and reconstructed tracks only~\cite{Wu:2026qlh}. There are also examples of multi-modal reconstruction models~\cite{Hart:2026dvw}, which use the principle of pre-training and fine-tuning without calling it a foundation model, as well as examples of how to use the foundation model paradigm for fine-tuning the full analysis chain for events~\cite{Vigl:2024lat}.

In this work, we show that the choice between single jets and full events need not be binary. OmniX operates on jet constituents as inputs and solves both classification and generative tasks at the level of individual jets and of complete events. This is achieved by training a decoder-only transformer jointly on next-token prediction and classification. The input consists of a concatenation of conditioning tokens, jet-level tokens, and constituent-level tokens. The features are binned individually and delayed in the spirit of SPADE~\cite{Birk:2026udp}. This allows the input to be binned finely enough to not lose resolution, and avoids having to train a separate tokenization model.

The rest of the paper is  structured as follows: 
We introduce the event-level and jet-level datasets in \cref{sec:datasets} and 
review the details of the OmniX architecture in \cref{sec:model}. We then show the results for generation and classification tasks in \cref{sec:results}, and close with a discussion in \cref{sec:conclusions}.

\section{Datasets}
\label{sec:datasets}

We use three simulated datasets (\cref{tab:datasets}): full QCD and \ttbar events from Collide-2V~\cite{Moreno:2025collide2v}, single jets from \jetclass~\cite{Qu:2022mxj,Qu:2022jetclass}, as well as the top-tagging
community dataset~\cite{Kasieczka:2019dbj,Kasieczka:2019toptagdata} for a downstream classification task. 
In the first two datasets, every particle is described by the same seven features: the transverse
momentum \pt, the pseudorapidity $\eta$, the azimuthal angle $\phi$, 
the electric charge $q$, the particle type, and the transverse and longitudinal impact parameters
$d_0$ and $d_z$. In the third dataset, only the four-momenta of the constituents are recorded.

\begin{table*}[t]
  \caption{Datasets used in this work. Split sizes are given in millions of jets or events.
  }
  \label{tab:datasets}
  \setlength{\tabcolsep}{10pt}
  \begin{tabular}{lllrrr}
    \toprule
    Dataset & Object & Classification target & Training & Validation & Test \\
    \midrule
    \multicolumn{6}{l}{\textit{Pre-training}} \\
    \jetclass & one AK8 jet & ten jet classes
      & \nJCTrain & \nJCVal & \nJCTest \\
    Collide-2V & full event, ${\leq}\,20$ \akfour jets & \akfour jet flavor ($b$, $c$, light) & & & \\
    \quad QCD inclusive & & & \nCvQCDTrain & \nCvQCDVal & \nCvQCDTest \\
    \quad QCD~$\to b\bar{b}$ & & & \nCvBBTrain & \nCvBBVal & \nCvBBTest \\
    \quad \ttbar, all hadronic & & & \nCvHadTrain & \nCvHadVal & \nCvHadTest \\
    \quad \ttbar, semi-leptonic & & & \nCvSemiTrain & \nCvSemiVal & \nCvSemiTest \\
    \quad \ttbar, all leptonic & & & \nCvLepTrain & \nCvLepVal & \nCvLepTest \\
    \midrule
    \multicolumn{6}{l}{\textit{Fine-tuning}} \\
    Top-tagging & one AK8 jet & top vs.\ QCD
      & \nTopTrain & \nTopVal & \nTopTest \\
    \bottomrule
  \end{tabular}
\end{table*}

\subsection{Collide-2V events}
\label{sec:data-c2v}
Collide-2V~\cite{Moreno:2025collide2v} is a simulated event-level dataset of 750 million events from 50 different Standard Model processes. The version we used\footnote{We obtained the dataset via direct communication with the authors. The 1\,M sample that is available at \texttt{huggingface.co/datasets/fastmachinelearning/collide-1m} differs from the final version we used.} is generated at $\sqrt{s}=\qty{14}{TeV}$ with MadGraph5\_aMC@NLO 3.5.8~\cite{Alwall:2014hca}, with hadronization and showering by Pythia 8.310~\cite{SciPostPhysCodeb.8, SciPostPhysCodeb.8-r8.3}. A modified version of Delphes, based on a pre-release of 3.5.1~\cite{deFavereau:2013fsa}, is used for detector simulation, with the CMS Phase-2 HL-LHC Delphes card. Pileup is added with $\langle\mu\rangle = 200$. The dataset contains a dual view of the same events: L1 trigger-level objects and fully reconstructed objects. In this work, only the fully reconstructed objects are used.

We use five processes: QCD inclusive ($H_\mathrm{T}>\qty{50}{GeV}$), QCD~$\to b\bar{b}$ ($H_\mathrm{T}>\qty{50}{GeV}$), \ttbar (all hadronic), \ttbar (semi-leptonic) and \ttbar (all leptonic). After removing events with event weight zero, we are left with \nCvTotal{} million events in total. We only use the pileup-per-particle identification (PUPPI)~\cite{Bertolini:2014bba,CMS:2020ebo} collections: all PUPPI particles as well as the \texttt{JetPuppiAK4} jets. 
Our preprocessing details can be found in \cref{app:c2v-data-processing}. After preprocessing, we select all \akfour jets with \pt > \qty{15}{GeV} (corresponding to the cut in Delphes) and $|\eta|<4$. No further kinematic cuts are made. The maximum number of \akfour jets per event with our selection is 20. In Collide-2V, the impact parameter information is stored without any smearing. To match JetClass, and to enable non-trivial flavor tagging, we add smearing to these variables using values extracted from Ref.~\cite{CERN-LHCC-2017-009}. This procedure is described in \cref{app:c2v-data-smearing}. After smearing, $d_0$ is clipped to $\pm\qty{25}{cm}$ and $d_z$ to $\pm\qty{60}{cm}$.

Finally, we split the dataset 80:10:10 into training, validation, and test splits.

\subsection{\jetclass}
\label{sec:data-jetclass}

\jetclass~\cite{Qu:2022mxj,Qu:2022jetclass} is a simulated dataset of \nJCTotal{} million jets in ten equally
populated classes, on which a large number of jet foundation models are trained (examples include ~\cite{Birk:2024knn,Golling:2024abg,Mikuni:2024qsr,Katel:2024ygn,Bardhan:2025icr,Hao:2025abk}). The hard processes are generated with
MadGraph5\_aMC@NLO~\cite{Alwall:2014hca}, and parton showering and hadronization are simulated with
Pythia~8.2~\cite{Sjostrand:2014zea}. No pileup is included. The detector response
is simulated with Delphes~\cite{deFavereau:2013fsa} in the CMS configuration. Jets are clustered from the
energy-flow objects with the \antikt algorithm~\cite{Cacciari:2008gp,Cacciari:2011ma} with distance parameter $R=0.8$ and kept if 
\begin{equation}
  \qty{500}{GeV} < p_{\mathrm{T},j} < \qty{1000}{GeV}, \qquad |\eta_j| < 2.
  \label{eq:sel-jetclass}
\end{equation}
One class contains jets initiated by light quarks or gluons, and the other nine contain jets from decays of
boosted Higgs bosons, top quarks, and $W$ and $Z$ bosons. A jet contains on average
\nConstJC{} constituents. Despite being AK8 jets, these jets have a small number of particles outside the jet cone. In order to remove noise, we
 remove the
\fracWindowJC\% of particles with $|\Delta\eta|>1.3$ or $|\Delta\phi|>1.3$ relative to the jet axis, all of them soft and charged. We do not add any further kinematic cuts, and use all jets in the official training, validation, and test splits.

\subsection{Top tagging}
\label{sec:data-toptag}

The top-tagging dataset~\cite{Kasieczka:2019dbj,Kasieczka:2019toptagdata} is a standard benchmark for
jet taggers and is also commonly used to evaluate pre-trained models (examples include~\cite{Qu:2022mxj,Katel:2024ygn,Bardhan:2025icr,Birk:2025fbs,Bhimji:2025isp}). Jets from
hadronically decaying top quarks (signal) and from light quarks and gluons (background) are generated with
Pythia~8~\cite{Sjostrand:2014zea} at $\sqrt{s}=\qty{14}{TeV}$. No pileup is included, and the detector response is simulated with Delphes~\cite{deFavereau:2013fsa}
in the ATLAS configuration. Jets are clustered from the energy-flow objects with the \antikt algorithm~\cite{Cacciari:2008gp,Cacciari:2011ma} with
distance parameter $R=0.8$. The leading jet of each event is kept if
\begin{equation}
  \qty{550}{GeV} < p_{\mathrm{T},j} < \qty{650}{GeV}, \qquad |\eta_j| < 2.
  \label{eq:sel-toptag}
\end{equation}
For each jet, only the four-momenta of up to 200 constituents are provided. Signal and background jets occur in equal proportions.

\begin{table*}[t]
  \caption{
  Codebooks of the particle features (top) and of the plan quantities (bottom). The
  number of bins excludes the padding and \texttt{missing} tokens. Symmetric-logarithmic bins with scale $s$ follow
  \cref{eq:symlog}, and quadratic bins are uniform in the square root of the underlying variable.
  Values outside the range are assigned to the
  first or last bin. For $d_0$ and $d_z$, the $+1$ is a separate bin for
  particles without a track, which are exactly the neutral particles.}
  \label{tab:binning}
  \begin{tabular}{llll}
    \toprule
    Quantity & Binning & Range & Bins \\
    \midrule
    \pt & logarithmic & \qtyrange{0.1}{6000}{GeV} & 1022 \\
    $\eta$, $\phi$ (constituents) & symmetric-log., $s=0.05$ & $\pm5$, $\pm\pi$ & 510 \\
    $\eta$, $\phi$ (out-of-jet) & linear & $\pm5$, $\pm\pi$ & 510 \\
    $d_0$ & symmetric-log., $s=\qty{0.005}{cm}$ & $\pm\qty{25}{cm}$ & $255+1$   \\
    $d_z$ & symmetric-log., $s=\qty{0.01}{cm}$ & $\pm\qty{60}{cm}$ & $255+1$   \\
    particle type, $j$ & categorical & -- & 8, 21 \\
    \midrule
    multiplicity $n$ & integer & 1--\nMaxPartCv\ (1--\nMaxPartJC) & \nMaxPartCv\ (\nMaxPartJC) \\
    jet \pt & logarithmic & \qtyrange{1}{6000}{GeV} & 3485 \\
    jet rapidity $y_j$ & linear & $\pm4$ & 4000 \\
    jet $\phi_j$ & linear & $\pm\pi$ & 3142 \\
    jet mass $m_j$ & quadratic & \qtyrange{0}{1000}{GeV} & 512 \\
    jet constituents $n_\mathrm{const}$ & integer & 0--255 & 256 \\
    \bottomrule
  \end{tabular}
\end{table*}

\section{OmniX}
\label{sec:model}

OmniX is a decoder-only transformer for single jets and full events. Like OmniJet-$\alpha$~\cite{Birk:2024knn},
one of its training tasks is next-token prediction, which requires discrete training targets. Instead of a learned tokenizer over the full phase space, such as a vector-quantized variational autoencoder
(VQ-VAE)~\cite{vandenOord:2017vqvae,Birk:2024knn,Golling:2024abg,Birk:2025wai,Birk:2025fbs} or its
residual-quantized variant (RQ-VAE)~\cite{Lee:2022rqvae,Krupa:2026ajb,Kakati:2026tca,Godi:2026kiy}, we adopt
the split-and-delay approach of SPADE~\cite{Birk:2026udp}, in which every feature of a particle is binned and tokenized on its own. This factorization keeps the total vocabulary size small, even with fine binning. 

\begin{figure*}[t]
  \centering
  \includegraphics[width=\linewidth]{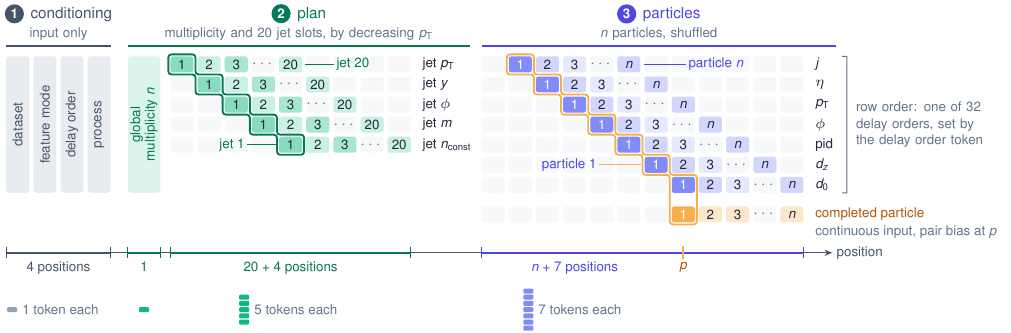}
  \caption{Sequence layout of OmniX, shown for one event. The sequence starts with conditioning tokens, indicating the dataset (jets or events), the feature mode (full features or kinematic features only), the delay order (controlling in which order the particle features are listed), and a process token (indicating the jet class or the event process type). The plan tokens include the particle multiplicity of the event, as well as jet-level features of up to twenty jets. The particle features include the features of all $n$ particles in the event, as well as their jet index $j$ indicating to which jet (if any) each particle belongs. The collected features of particle~1 are outlined along the diagonal, shown in one of the 32 delay orders. These form a completed particle at position $p$, which lets the model understand the full particle despite the delay.}
  \label{fig:architecture-sequence}
\end{figure*}

\Cref{fig:architecture-sequence} illustrates the split-and-delay mechanism of SPADE, and \cref{fig:architecture-network} shows the architecture of the model. The full pipeline is discussed in the following.

\subsection{Particle properties and discretization}
\label{sec:encoding}

Every particle carries the seven features: \pt, $\eta$, $\phi$, the electric charge $q$,
the particle type, $d_0$, and $d_z$. We merge $q$ and the particle type into one categorical feature, pid, with
the eight categories $h^\pm$, $h^0$, $\gamma$, $e^\pm$, and $\mu^\pm$. To these six features, we
add the jet index $j\in\{0,\dots,20\}$, which either assigns the particle to one of at most twenty jets, selected
as in \cref{sec:data-jetclass,sec:data-c2v} and numbered by decreasing \pt ($j\geq1$), or marks it as an
out-of-jet particle ($j=0$). The constituents of a jet
carry $\eta$ and $\phi$ as offsets $\eta-y_j$ and $\phi-\phi_j$ from the jet axis, where $y_j$ and $\phi_j$ are
the rapidity and the azimuth of the four-momentum sum of its constituents, each taken as massless. Out-of-jet particles keep absolute coordinates, after the event has been rotated in azimuth such that
the leading jet lies at $\phi=0$. If an event has no selected jet, its particle with the highest \pt is rotated
to $\phi=0$ instead. In this representation, an
event consists of up to twenty jets and the particles outside them. Single jets from \jetclass are events with exactly one jet and no particles outside it. All jets are discretized with the same codebooks,
one for each of the seven features (\cref{tab:binning}, top).

All datasets share the same codebooks, including their binning and range. The binning type of each
codebook follows the shape of the distribution of its feature. The relative $\eta$ and $\phi$ of the constituents, as well as the trajectory displacement features 
$d_0$ and $d_z$, are signed and peak at zero. They are therefore binned
uniformly in
\begin{equation}
  u=\operatorname{sgn}(x)\,\ln\!\left(1+\frac{|x|}{s}\right),
  \label{eq:symlog}
\end{equation}
a symmetric-logarithmic transform, where $x$ is the value of the feature and $s$ is a scale that is set per feature (listed in \cref{tab:binning}). The bins are fine for
$|x|\lesssim s$ and increasingly coarse above it. The absolute $\eta$ and $\phi$ of out-of-jet
particles are binned uniformly, with the same number of bins and thus the same tokens. Which of the two binnings
applies is fixed by the jet index.
In \cref{app:discretization}, we show that encoding and decoding the test data leaves all fidelity
metrics unchanged within their statistical precision.

\subsection{Sequence layout}
\label{sec:sequence}
Next-token prediction requires every event to be written as a sequence. \Cref{fig:architecture-sequence} shows
this sequence, which we describe from left to right.

Since OmniX models jets and events from different datasets, the sequence begins with conditioning
tokens, which state what kind of event follows. They are given as input and never predicted. First, the dataset token states whether the event is from \jetclass or from Collide-2V. The datasets also differ in the particle features they provide. The top-tagging dataset, for instance, provides only the four-momenta of the particles. To present such
data in a form the model has seen in training, we add a kinematic mode, which the feature-mode token
selects. The feature mode of a training event is drawn at random, full or kinematic with equal probability.
In the kinematic mode, pid, $d_0$, and $d_z$ are replaced by the \texttt{missing} token and excluded from
the loss, so that the model also learns to describe the particles by their kinematics alone. A jet from the top-tagging dataset then has the same form as a \jetclass jet in
the kinematic mode. To generate events of a chosen kind, the process token can name a jet class or a Collide-2V process.

Before the first particle, the model
predicts a plan, which describes the event coarsely by its multiplicity and its jets. The event is thus generated
from coarse to fine, with the particles filling in the details after the plan. Both the plan and the particles enter the generative loss. The plan begins with the particle multiplicity $n$. This fixes the number of particles during generation,
so that no end-of-sequence token is needed. Twenty jet slots follow, each describing one jet by its \pt, its axis $(y_j,\phi_j)$, and its mass $m_j$, together with its number of constituents $n_\mathrm{const}$. Each quantity has its own codebook
(\cref{tab:binning}, bottom). Unused slots are filled with zeros. As in SPADE~\cite{Birk:2026udp}, each of the
five slot quantities fills its own row of the sequence, delayed by one position against the previous one, so that the
twenty slots occupy 24 positions. 

The particles of an event form a point cloud without a natural order. Rather than imposing an arbitrary order for
the next-token prediction task, we shuffle the particles every time an event is seen in training. The model is thus trained to describe an unordered set, even though the network itself is not permutation
invariant. Just like the elements of the plan, the particles are arranged in the split-and-delay layout, with one row for each of their seven features, so that the collected features of a particle lie on a diagonal (as outlined for particle~1 in \cref{fig:architecture-sequence}). The benefit of this layout is that it keeps the sequence short. A flattened version would require a sequence of length $N = 7n$, which scales badly in the attention mechanism ($\mathcal{O}(N^2)$). With this layout, the $n$ particles of an event instead occupy only $N = n+7$ positions. 

The elements of the sequence differ in their number of tokens: a conditioning element has one, a jet of the plan five, and a particle seven. Each position must nevertheless enter the network as one vector of the model dimension $h$. Conditioning tokens and the jet quantities of the plan are embedded directly as vectors of size $h$, and the jet quantities that share a position are summed. For the particles, each of the seven tokens is first embedded into a smaller vector. The seven vectors are then concatenated and projected to $h$ by a linear layer.  In this way, all elements in the sequence, regardless of whether they have one token (conditioning features), five tokens (jet-level plan) or seven tokens (particles), are mapped to the same embedding dimension. The delay itself makes sure that the inter-feature correlations are preserved. At position $p$, feature $i$ of a particle will be predicted based on the previous $i-1$ features of this particle.

With a fixed order of the rows, every feature would always be predicted conditioned on the same set of earlier features. OmniX therefore varies the order from event to event. Two rules restrict the possible orders. The jet index is always placed before the kinematics, because $\eta$ and $\phi$ of a jet constituent are given relative to the axis of its jet. Since only charged particles leave a track with $d_0$ and $d_z$, pid is always placed before them. Of the 420 orders that satisfy both rules, a fixed random subset of 32 is used. One of them is drawn for every event and given to the model as the delay-order token at the start of the sequence.

\begin{figure*}[t]
  \centering
  \includegraphics[width=\linewidth]{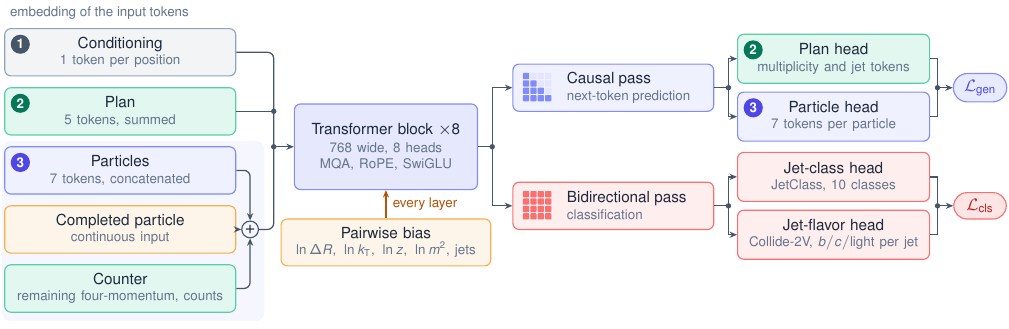}
  \caption{The network of OmniX and the two passes of a training step. It runs twice per training step with
  shared weights: causally for the generative loss and bidirectionally for the classification loss. Colors follow \cref{fig:architecture-sequence}, with the counter in the color of the plan it is computed from, and red marks classification. The inputs are drawn once per type of position, numbered as in \cref{fig:architecture-sequence}: conditioning~(1), plan~(2), and particles~(3). The particle token embeddings are combined with the continuous input of the completed particles (i.e., without the delay), as well as with the counter vector that keeps track of the remaining four-momentum of the jet (according to the plan). Their sum then enters a transformer decoder, where pairwise bias vectors are added at every layer. The output from the transformer goes to four heads: a plan head and a particle head, which together form the generative head, a jet classification head (for JetClass) and a jet flavor tagging head (for jets from Collide-2V).}
  \label{fig:architecture-network}
\end{figure*}

\subsection{Tokens, particles, counters}
\label{sec:inputs-outputs}

At every particle position, the network embeds the token of each row with the dedicated codebook, concatenates the seven embeddings, and
projects them to the model width (\cref{fig:architecture-network}). Since all targets are discrete, it predicts
the token of every row at the next position as a distribution over the bins of its codebook. The generative loss
$\mathcal{L}_\text{gen}$ is the sum of the cross-entropies of the seven particle features and the six plan
quantities, each averaged over all its tokens in a batch. Padding and \texttt{missing} tokens are excluded,
whereas the plan tokens of unused jet slots are included.

The pid of a particle fixes which bins of $d_0$ and $d_z$ are possible. A neutral particle always takes the bin
for particles without impact parameters, and a charged particle never does. Since pid precedes $d_0$ and $d_z$
in every delay order, their predicted distributions are restricted to the bins that the pid allows. The
restriction applies in training and in generation, so it is part of the likelihood, and no generated particle
breaks the rule.

Because of the delay, no position holds all features of one particle. At the position
$p$ where a particle is completed, an MLP therefore embeds its \pt, $\eta$, and $\phi$ together as numbers and
adds the result to the token embedding. In generation, the model samples only a bin and not the value within it.
If the network read exact values in training, it would learn to rely on a precision that generation cannot
provide. In training and in generation alike, the value is thus drawn at random within the bin of its token
(dithered), uniformly or, for the logarithmic \pt bins, log-uniformly.

Since the particles are shuffled, the constituents of a jet that are already in the sequence are scattered over
all earlier positions. To know how much of a planned jet is still missing, the network would have to find these
constituents, count them, and sum their four-momenta. Instead, a special counter (one per jet) provides these numbers directly. Once the jet index has been predicted, the correct counter can be called. It 
keeps track of the remaining four-momentum, how many constituents the jet has so far and how many are still to come according to the plan. If the jet index is zero (i.e., the next particle lies outside the jets), the counter is zero. An MLP maps the counter to the model width, and the result is added to the embedding at every
position. The values in the counter are recomputed at every position.

\subsection{Pairwise attention bias}
\label{sec:pairwise}
As in ParT~\cite{Qu:2022mxj}, an additive bias on the attention logits gives the attention direct access to the kinematics of every pair of particles. For two particles completed at positions $p$ and $q$, the bias uses their angular distance $\Delta R$, the relative transverse momentum $k_\mathrm{T}=\min(p_{\mathrm{T},p},p_{\mathrm{T},q})\Delta R$, the fraction $z=\min(p_{\mathrm{T},p},p_{\mathrm{T},q})/(p_{\mathrm{T},p}+p_{\mathrm{T},q})$ of the summed \pt that the softer particle carries, and the squared pair mass $m^2$, all evaluated for massless particles. All four enter as logarithms. Jet constituents store $\eta$ and $\phi$ relative to the axis of their own jet, which cannot be compared between particles of different jets. The pair features are therefore computed in absolute coordinates, obtained by adding the planned jet axis back to every constituent, so that pairs within a jet, across jets, and with out-of-jet particles are described in one frame. Flags mark whether the two particles share a jet and whether one or both lie outside all jets. An MLP maps the features and flags to one bias per attention head, which is added in all eight layers. Since only completed particles enter, the bias between $p$ and an earlier position $q$ uses no token after $p$ and respects the causal mask.

\subsection{Joint generation and classification}
\label{sec:joint}

OmniX is trained to generate and to classify at the same time, with two passes through the full backbone in
every training step (\cref{fig:architecture-network}). The generative pass is causal, so that every token is
predicted from all earlier tokens. The classification pass, on the other hand, is bidirectional, so that every position can attend to
the full event. The two passes share all weights except their task heads. The total loss
is the sum of their losses, $\mathcal{L}=\mathcal{L}_\text{gen}+\mathcal{L}_\text{cls}$. 

The process token and the classification target depend on the dataset. For \jetclass, the process token holds one of the ten jet classes, and the classifier predicts this class. For Collide-2V, the process token holds one of the five processes. The classifier, however, does not predict the process but the flavor\footnote{Since the current model is trained on QCD and \ttbar processes only, event classification is too easy and is left out for now.} ($b$, $c$, or light) of every \akfour jet. The jet flavor itself never enters as a token.

The process token is set differently in the generative and the classification pass. In the generative pass, it is set to \texttt{NULL} with
probability 0.5. The model thus learns a conditional and an unconditional generator at the same time, with the
conditioning dropout of classifier-free guidance~\cite{Ho:2022cfg}. In generation, the process token can therefore be set to a jet class or a process, or left at \texttt{NULL}. In the classification pass, the process token is always set to \texttt{NULL} to prevent label leakage. The plan and all other tokens stay visible, so the classifier sees the full event, including its planned jets.

The generative head is one linear layer per particle feature and per plan quantity. The classification heads turn the particles of a jet into a label.

A \jetclass jet has one label. We append one learned class token to the end of the particle sequence. Two class-attention blocks~\cite{Touvron:2021cait,Qu:2022mxj} let this token look at all particles of the jet, and an MLP maps it to the jet class. A Collide-2V event has one flavor label per \akfour jet. We therefore use one class token per jet. Each class token is initialized with an embedding of the \pt, the axis, and the number of constituents of its jet, so that it knows which jet it stands for. Three transformer blocks then let the class tokens look at the particles, where a learned bias points each token to the particles of its own jet, and at each other. The jets of an event are thus not tagged in isolation. Each jet is tagged with access to the full event and to what the other class tokens have gathered about their own jets, so that evidence for the flavor of one jet can inform the flavor of another. An MLP finally maps each class token to the flavor of its jet.

\subsection{Architecture and training}
\label{sec:pretraining}

The backbone is a decoder-only transformer of eight pre-layer-norm blocks~\cite{Xiong:2020preln} of width 768,
with multi-query attention~\cite{Shazeer:2019mqa} with eight heads, rotary position
embeddings~\cite{Su:2021rope}, SwiGLU feed-forward layers~\cite{Shazeer:2020glu} of hidden size 3072, and no
dropout. The backbone has \nParBackbone~million parameters.

We train three models with this backbone, which differ in their pre-training data: 
\jetclass jets alone (Jets-only), 
Collide-2V events alone (Events-only), and \jetclass jets and Collide-2V events together (Events+Jets), listed in \cref{tab:models}. Every batch is drawn from a single dataset, in the ratio given in \cref{tab:models}. In
Events+Jets, each batch therefore trains only the classification head of its own dataset.

\begin{table}[t]
  \caption{The three pre-trained models. The batch ratio counts batches from \jetclass and from the Collide-2V
  processes QCD inclusive, QCD~$\to b\bar{b}$, and \ttbar all hadronic, semi-leptonic, and all leptonic. This ratio follows the number of training events in each dataset, such that the number of full passes over each process becomes approximately equal. Parameters include all
  embeddings and heads. The constant phase at the peak learning rate and the final step are given in optimizer
  steps. Dataset passes are counted over the training split, separately for \jetclass and Collide-2V in
  Events+Jets. GPU-hours include start-up and validation.}
  \label{tab:models}
  \begin{tabular}{@{}llll@{}}
    \toprule
     & Events+Jets & Jets-only & Events-only \\
    \midrule
    Data & \jetclass, Collide-2V & \jetclass & Collide-2V \\
    Batch ratio & 50:5:1:2:2:2 & -- & 0:5:1:2:2:2 \\
    Parameters (M) & \nParUnion & \nParJets & \nParEvents \\
    Constant phase & \stepsConstUnion & \stepsConstJets & \stepsConstEvents \\
    Final step & \stepsFinalUnion & \stepsFinalJets & \stepsFinalEvents \\
    Dataset passes & \passesUnionJC, \passesUnionCv & \passesJets & \passesEvents \\
    H100 GPU-hours & \gpuhUnion & \gpuhJets & \gpuhEvents \\
    \bottomrule
  \end{tabular}
\end{table}

All models are trained with the Ranger optimizer, which combines RAdam~\cite{Liu:2020radam} with
Lookahead~\cite{Zhang:2019lookahead}, on batches of 1024 events. After a linear warm-up, the learning rate is held
constant at $10^{-3}$ for at most \num{2457600} steps. We start the anneal over a further 20\% of the steps once the validation loss shows
diminishing returns. More information about the model and hyperparameters is available in \cref{app:hyperparameters}.

\begin{figure*}[t]
  \centering
  \includegraphics[width=\linewidth]{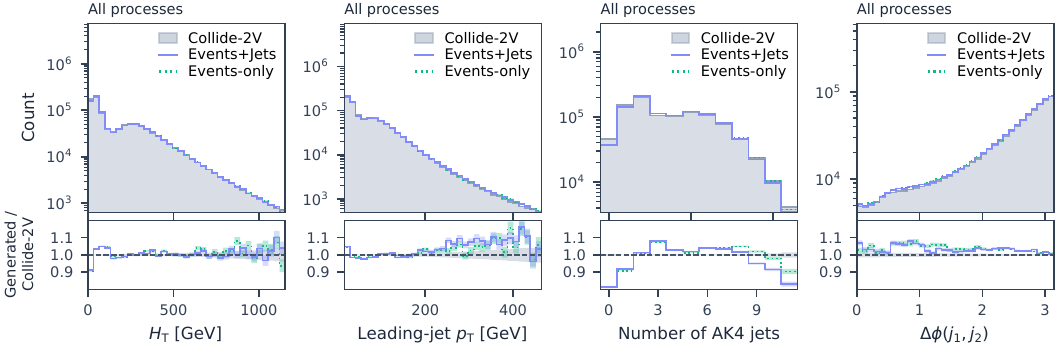}
  \caption{
  Distributions of event observables for Collide-2V events generated by the Events+Jets and the Events-only model with the process token given. Each sample, the test set and the generated events of either model, contains 1\,M events with the five processes in the same proportions as in the training data. $H_\mathrm{T}$ is the scalar sum of the transverse momenta of the AK4 jets, and $\Delta\phi(j_1,j_2)$ is the azimuthal angle between the two leading AK4 jets. In all events, generated and from the test set, the jets are reclustered from all particles in the event (anti-$k_\mathrm{T}$, $R=0.4$, $\pt>\qty{15}{GeV}$, $|\eta|<4$). The lower panels show the ratio to the Collide-2V test events, with their statistical uncertainty as a gray band. Further observables are shown in \cref{fig:app-c2v-tt,fig:app-c2v-qcd} in \cref{app:fidelity}.
  }
  \label{fig:events-panels}
\end{figure*}

\subsection{Generation}
\label{sec:generation}

To generate, the conditioning tokens are set: the dataset, the feature mode, the delay order, and optionally the process token (the jet class for \jetclass, the process for
Collide-2V events). The plan and the following particles are
sampled token by token with a key-value cache, at temperature one. Continuous features are decoded to values
by dithering (\cref{sec:inputs-outputs}), whereas categorical tokens and the bin for particles
without impact parameters are decoded exactly.

\section{Results}
\label{sec:results}

We compare the three models of \cref{tab:models}, which share the architecture and the training setup and differ only in their training data: Events+Jets, Events-only, and Jets-only. We first compare the generated events and jets in \cref{sec:res-generation}. We then evaluate the classification after pre-training, for \jetclass jet tagging in \cref{sec:res-jetclass} and for flavor tagging in Collide-2V events in \cref{sec:res-flavor}. Finally, we fine-tune the models on the top-tagging dataset in \cref{sec:res-finetune}.

\begin{figure*}[t]
  \centering
  \includegraphics[width=\linewidth]{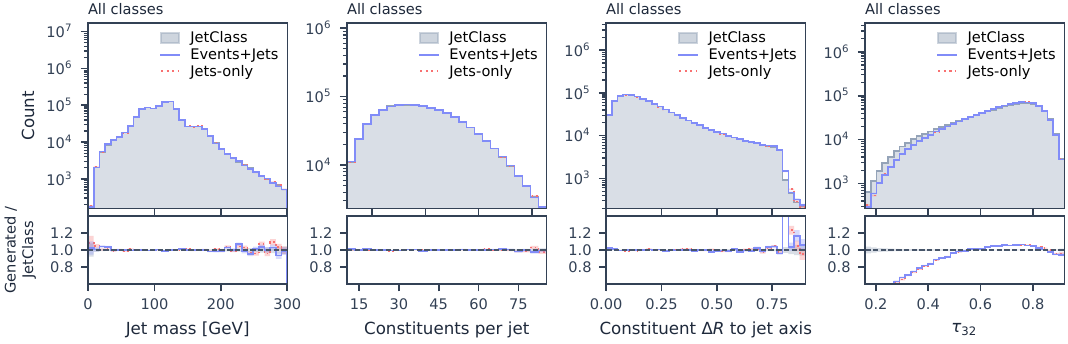}
  \caption{Distributions of jet observables for \jetclass jets generated with the process token set to the jet class by the Events+Jets and the Jets-only model (1\,M each, 100\,k per class), compared to the test jets of all ten classes. $\tau_{32}=\tau_3/\tau_2$ is the ratio of the
  $N$-subjettiness values~\cite{Thaler:2010tr}, computed with winner-take-all $k_\mathrm{T}$ axes and $\beta=1$. The lower panels show the
  ratio to the JetClass test jets, with the statistical uncertainty of the test jets as a gray band. Further observables are shown in
  \cref{fig:app-jets-jet} in \cref{app:fidelity}.}
  \label{fig:jets-panels}
\end{figure*}

\subsection{Generation}
\label{sec:res-generation}

\subsubsection{Full events}
\label{sec:res-events}
\Cref{fig:events-panels} compares the generated events with the Collide-2V test events in four observables at the event and jet level. Both the Events+Jets and the Events-only model follow the Collide-2V distributions closely, and their curves lie almost on top of each other in all four panels. Neither training set shows a clear advantage.

The deviations are small and similar in both models. In $H_\mathrm{T}$, both agree with the test events closely from 130\,GeV up to the statistically limited tail near 1\,TeV, and generate about 5\% too many events between 30 and 130\,GeV. The leading-jet \pt agrees best with the test events between 30 and 200\,GeV, and both models show similar deviations in the high-\pt tail. The largest deviation appears in the number of jets, where both models generate around 20\% too few events without a jet, 10\% too few with one jet, and up to 10\% too many with three jets. As a result, more of the generated events than test events have at least two jets, which appears as an offset of a few percent in $\Delta\phi(j_1,j_2)$.
The models also assign a jet index to every generated particle. When the jets are built from this index instead of being reclustered, these deviations are not observed. This comparison is shown in \cref{fig:app-c2v-modeljets} in \cref{app:fidelity}.

\Cref{app:fidelity} shows further event, jet, and particle-level observables in \cref{fig:app-c2v-tt,fig:app-c2v-qcd}, separately for the \ttbar and the QCD processes, together with distance metrics between the generated and the test events. In these metrics as well, both models are similarly close to the test events.

\subsubsection{Jets}
\label{sec:res-jets}

\Cref{fig:jets-panels} compares the generated jets with the \jetclass test jets in four observables at the jet and constituent level. The Events+Jets and the Jets-only model both
follow the \jetclass distributions closely, and their curves lie on top of each other in all four panels. Neither training set shows a clear advantage. 

The jet mass, the number of constituents, and $\Delta R$ agree with the test jets over nearly the full range, with small deviations only at the edges of the distributions. The one clear deviation appears in $\tau_{32}$, where both models generate too few jets with low $\tau_{32}$ and slightly too many between 0.6 and 0.8. In the test set, the jets with low $\tau_{32}$ come mainly from top-quark and $H\to\ell\nu qq'$ decays. The deficit is identical for both models and thus independent of the training data. We attribute it to the autoregressive generation, in which the positions of the three prongs emerge only gradually and early particles placed off their prong axis wash out the three-prong structure.

\Cref{app:fidelity} shows further jet and constituent observables in \cref{fig:app-jets-jet}, together with
distance metrics between the generated and the test jets in \cref{tab:metrics-test-jets}. In these metrics, the
two models are close to each other, with small differences in favor of the Events+Jets model.

\begin{table}[t]
  \centering
  \caption{Classification performance on the \jetclass test set for the Events+Jets and the Jets-only model. We list the accuracy, the one-vs-one AUC, and the QCD rejection for each signal class at the signal efficiency given in parentheses. For the accuracy and the AUC
  the statistical uncertainties are below $10^{-4}$ and not shown.}
  \label{tab:jetclass-parity}
  \begin{tabular*}{\linewidth}{@{}l@{\extracolsep{\fill}}r@{\extracolsep{0pt}${}\pm{}$}l@{\extracolsep{\fill}}r@{\extracolsep{0pt}${}\pm{}$}l@{}}
\toprule
 & \multicolumn{2}{c}{Events+Jets} & \multicolumn{2}{c}{Jets-only} \\
\midrule
Accuracy & \multicolumn{1}{r}{0.8514} &  & \multicolumn{1}{r}{0.8505} &  \\
AUC & \multicolumn{1}{r}{0.9862} &  & \multicolumn{1}{r}{0.9861} &  \\
\midrule
\multicolumn{5}{@{}l}{QCD rejection ($\epsilon_S$)} \\
$H\to b\bar{b}$ (50\%) & 10100 & 700 & 10200 & 700 \\
$H\to c\bar{c}$ (50\%) & 3110 & 120 & 3070 & 120 \\
$H\to gg$ (50\%) & 109 & 1 & 108 & 1 \\
$H\to 4q$ (50\%) & 1247 & 31 & 1214 & 30 \\
$H\to \ell\nu qq'$ (99\%) & 4590 & 220 & 4510 & 210 \\
$t\to bqq'$ (50\%) & 20600 & 2100 & 21500 & 2200 \\
$t\to b\ell\nu$ (99.5\%) & 18000 & 1700 & 17700 & 1700 \\
$W\to qq'$ (50\%) & 392 & 5 & 390 & 5 \\
$Z\to q\bar{q}$ (50\%) & 315 & 4 & 313 & 4 \\
\bottomrule
\end{tabular*}

\end{table}

\subsection{\jetclass classification}
\label{sec:res-jetclass}

Both the Events+Jets and the Jets-only model are trained with a classification objective on the ten \jetclass classes, so we can evaluate them directly after pre-training without fine-tuning. We classify the 20\,M jets of the \jetclass test set with the Events+Jets and the Jets-only model, using the exact values of the continuous inputs instead of the binned and dithered ones. \Cref{tab:jetclass-parity} lists the accuracy of the ten-class prediction, the one-vs-one AUC, and the QCD rejection of each of the nine signal classes at a fixed signal efficiency. For the rejection of a signal class $S$, we use the discriminant $p_S/(p_S+p_\text{QCD})$ and quote the inverse of the QCD efficiency.

The two models perform equally well. Their accuracies of 0.8514 (Events+Jets) and 0.8505 (Jets-only) are slightly below those of dedicated \jetclass classifiers such as ParT~\cite{Qu:2022mxj}, which reaches 0.861. Between the two OmniX models, the accuracy differs by 0.0009 and the AUC by 0.0001, and for each of the nine signal classes the QCD rejections agree within their statistical uncertainties. The Events+Jets model is nominally higher in accuracy, AUC, and seven of the nine rejections, but with one training per model we cannot separate such differences from the variation between trainings. Additional training on Collide-2V events therefore neither improves nor degrades the classification of \jetclass jets.

\begin{figure*}[t]
  \centering
  \includegraphics[width=\linewidth]{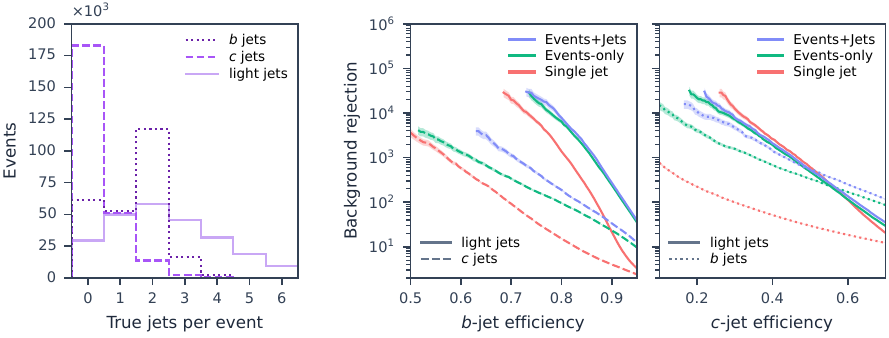}
  \caption{Jet flavor tagging on the labeled \akfour jets of the Collide-2V test events. Left: number of $b$, $c$, and light jets per event. Middle and right: background rejection $1/\epsilon_B$ as a function of the signal efficiency for $b$ tagging and $c$ tagging, for the Events+Jets and the Events-only model, as well as for the Single jet model. The line style indicates the background flavor. Jets are ranked by the discriminant $p_S/(p_S+p_B)$, built from the predicted probabilities of the signal flavor and of the background flavor. Bands show the statistical uncertainty, and curves end where fewer than 20 background jets pass the selection.}
  \label{fig:flavor-tagging}
\end{figure*}

\subsection{Jet flavor tagging}
\label{sec:res-flavor}

\Cref{fig:flavor-tagging} shows the OmniX models as $b$ and $c$ taggers on the \akfour jets of the
Collide-2V test events. The left panel gives the number of $b$, $c$, and light jets per event. Since $b$
jets appear mostly in pairs, a tagger that sees the whole event can use the other jets as additional
information during tagging. To measure this effect, we add a third model, which we call Single jet. It is the Jets-only
model fine-tuned on single \akfour jets of Collide-2V, and serves as our jet-level reference in which each
jet is tagged individually without the rest of the event.

\begin{table}[t]
  \centering
  \caption{Background rejection $1/\epsilon_B$ of the jet-flavor heads on all labeled \akfour jets of the Collide-2V test events, for the Events+Jets, the Events-only, and the Single jet model. The $b$ tagger is evaluated at a $b$-jet efficiency of 70\%, the $c$ tagger at a $c$-jet efficiency of 30\%, and each column gives the rejection of one background flavor. Where fewer than 20 background jets pass the working point, the rejection is limited by the size of the test set, and we give the lower limit at 95\% confidence level. Bold marks the highest rejection where it exceeds the second highest by more than two standard deviations.}
  \label{tab:flavor-wp}
  \footnotesize
\begin{tabular*}{\linewidth}{@{}l@{\extracolsep{\fill}}r@{\extracolsep{0pt}\,}l@{\extracolsep{\fill}}r@{\extracolsep{0pt}\,}l@{\extracolsep{\fill}}r@{\extracolsep{0pt}\,}l@{\extracolsep{\fill}}r@{\extracolsep{0pt}\,}l@{}}
\toprule
 & \multicolumn{4}{c}{$b$ tagging} & \multicolumn{4}{c}{$c$ tagging} \\
 & \multicolumn{4}{c}{$\epsilon_b=70\%$} & \multicolumn{4}{c}{$\epsilon_c=30\%$} \\
\cmidrule(lr){2-5}\cmidrule(lr){6-9}
Background & \multicolumn{2}{c}{$c$ jets} & \multicolumn{2}{c}{Light jets} & \multicolumn{2}{c}{$b$ jets} & \multicolumn{2}{c}{Light jets} \\
\midrule
Events+Jets & \bfseries 1130 & \bfseries\boldmath $\pm$\,130 & \multicolumn{2}{c}{$>34000$} & \bfseries 4700 & \bfseries\boldmath $\pm$\,500 & 7900 & $\pm$\,900 \\
Events-only & 337 & $\pm$\,21 & \multicolumn{2}{c}{$>47000$} & 1580 & $\pm$\,110 & 7300 & $\pm$\,800 \\
Single jet & 92 & $\pm$\,3 & 19700 & $\pm$\,3500 & 100 & $\pm$\,2 & \bfseries 12600 & \bfseries\boldmath $\pm$\,1800 \\
\bottomrule
\end{tabular*}

\end{table}

The middle panel shows $b$ tagging. Both models with access to the full event reject $c$ jets and light
jets better than the Single jet model, and the Events+Jets model rejects $c$ jets best.
\Cref{tab:flavor-wp} gives the rejections at a $b$-jet efficiency of 70\%. There, the Events-only model
rejects $c$ jets about four times better than the Single jet model, and the Events+Jets model another
three times better than the Events-only model. For both event models, so few light jets pass that we can
only give a lower limit on the light-jet rejection.

The right panel shows $c$ tagging. Against $b$ jets, the order of the three models is the same and the
differences are larger than for the $c$-jet rejection in $b$ tagging. At a $c$-jet efficiency of 30\%, the Events-only model rejects $b$ jets about
sixteen times better than the Single jet model, and the Events+Jets model again three times better than
the Events-only model. Against light jets, these advantages disappear. The two event models agree within
their uncertainties, and the Single jet model is slightly ahead at the working point.

The event models thus gain where $b$ jets have to be identified, in line with the pairs seen in the left
panel. The additional gain of the Events+Jets model does not come from more flavor-labeled events. Both
event models learn the jet flavor from the same Collide-2V labels, and the Events+Jets model passed over
the Collide-2V training events \passesUnionCv{} times, compared with \passesEvents{} for the Events-only
model. We therefore attribute it to the \jetclass jets in the training set, possibly to the classes
$H\to b\bar{b}$ and $H\to c\bar{c}$,  which require the model to distinguish decays to $b$ quarks from
decays to $c$ quarks. Each model was trained once, so the uncertainties do not include the variation between trainings.

\begin{figure}[t]
  \centering
  \includegraphics[width=\linewidth]{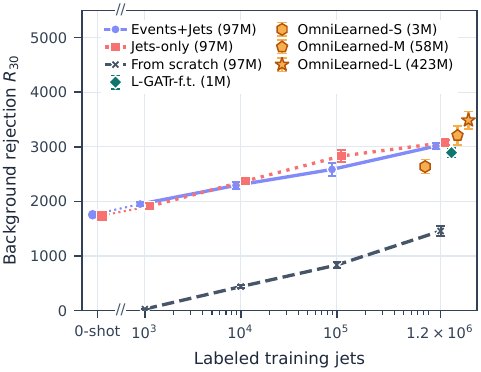}
  \caption{Background rejection $R_{30}=1/\epsilon_B$ at a signal efficiency of $\epsilon_S=0.3$ on the
  top-tagging dataset as a function of the number of labeled training jets. The zero-shot points show the
  pre-trained models evaluated without any fine-tuning. The from-scratch network is the same network trained from
  random initialization on the labeled jets alone. Error bars show the standard deviation over five fine-tunings. Published results for OmniLearned~\cite{Bhimji:2025isp} and the fine-tuned
  L-GATr~\cite{Brehmer:2024yqw} on the full training set are shown with their number of parameters and the
  uncertainties quoted in their papers. Points at the same training set size are offset horizontally for
  readability.}
  \label{fig:toptag-ladder}
\end{figure}

\begin{table*}[t]
  \centering
  \caption{Top tagging with the full training set. $R_{50}$ and $R_{30}$ are the background rejections
  $1/\epsilon_B$ at signal efficiencies $\epsilon_S$ of 0.5 and 0.3. Uncertainties are the standard deviation over five fine-tunings. Where quoted, they are at most $2\times10^{-4}$ for accuracy and AUC and are not shown. Reference
  values are taken from their papers. Model parameters are those of the fine-tuned network. Parameters and pre-training set sizes are rounded to millions.}
  \label{tab:toptag}
  \sisetup{group-digits=none}
\begin{tabular*}{\linewidth}{@{\extracolsep{\fill}}l r r@{\extracolsep{0pt}}l@{\extracolsep{\fill}} S[table-format=1.3] S[table-format=1.4] S[table-format=3.0]@{\extracolsep{0pt}${}\pm{}$}S[table-format=2.0]@{\extracolsep{\fill}} S[table-format=4.0]@{\extracolsep{0pt}${}\pm{}$}S[table-format=3.0]@{}}
\toprule
 & Model parameters & \multicolumn{2}{l}{Pre-training data} & {Accuracy} & {AUC} & \multicolumn{2}{c}{$R_{50}$} & \multicolumn{2}{c}{$R_{30}$} \\
\midrule
Events+Jets & 97\,M & 100\,M & ~jets + 23\,M events & 0.944 & 0.9879 & 679 & 9 & 3014 & 56 \\
Jets-only & 97\,M & 100\,M & ~jets & 0.944 & 0.9879 & 641 & 14 & 3079 & 49 \\
\midrule
OmniLearned-S~\cite{Bhimji:2025isp} & 3\,M & 1058\,M & ~jets & 0.944 & 0.9875 & 565 & 12 & 2637 & 128 \\
OmniLearned-M~\cite{Bhimji:2025isp} & 58\,M & 1058\,M & ~jets & 0.944 & 0.9880 & 656 & 12 & 3208 & 176 \\
OmniLearned-L~\cite{Bhimji:2025isp} & 423\,M & 1058\,M & ~jets & 0.944 & 0.9880 & 688 & 9 & 3486 & 157 \\
L-GATr-f.t.~\cite{Brehmer:2024yqw} & 1\,M & 100\,M & ~jets & 0.945 & 0.9879 & 651 & 11 & 2894 & 84 \\
\bottomrule
\end{tabular*}

\end{table*}

\subsection{Fine-tuning on top tagging}
\label{sec:res-finetune}

In the last test we measure how well the pre-trained models adapt to a new dataset. We fine-tune the Events+Jets and the Jets-only model on the top-tagging dataset described in \cref{sec:data-toptag}, a benchmark that is
commonly used to evaluate pre-trained jet models. Both models have already learned to separate top-quark jets
from QCD jets in \jetclass. The top-tagging dataset nevertheless differs from \jetclass in its detector
simulation and its jet selection, and it provides only the four-momenta of the constituents.

As described in \cref{sec:sequence}, the jets are therefore presented with the \jetclass dataset token and in
the kinematic mode. The top-tagging head is the pre-trained jet-class head, of which we keep only the two outputs
for QCD and $t\to bqq'$ jets. We train with a cross-entropy loss on 1\,k, 10\,k, 100\,k, and finally all
1.2\,M labeled training jets. For each training set size we run five fine-tunings with different seeds,
select the checkpoint with the highest validation AUC, and evaluate it on the full test set. We also evaluate both pre-trained models on the test set without any fine-tuning, which we call zero-shot. As a baseline, the same network is trained from random initialization on the labeled jets alone (referred to as \textit{from scratch}). \Cref{tab:hyper-finetune} in
\cref{app:hyperparameters} lists all settings.

\Cref{fig:toptag-ladder} shows the background rejection $R_{30}$, the inverse of the background efficiency at a signal efficiency of 30\%, as a function of the number of labeled training jets. Without any fine-tuning, both pre-trained models already reject more background than the from-scratch network trained on all 1.2\,M jets. With 1\,k labeled jets, the from-scratch network barely separates signal from background, while the fine-tuned models improve on their zero-shot rejection. Their rejection then rises with every increase of the training set. The two pre-trained models lie close together at every training set size, with differences
comparable to the spread of the five fine-tunings.

\Cref{tab:toptag} compares the models fine-tuned on the full training set with recent pre-trained taggers. These are OmniLearned~\cite{Bhimji:2025isp} in three sizes, pre-trained on about one billion jets, and the fine-tuned L-GATr~\cite{Brehmer:2024yqw}, a Lorentz-equivariant network with about one million parameters pre-trained on \jetclass. Accuracy and AUC are nearly identical for all models, so the comparison rests on the background rejections. In $R_{30}$, both OmniX models clearly exceed OmniLearned-S, are compatible with OmniLearned-M, and reject at least as much background as the fine-tuned L-GATr. Only OmniLearned-L, with four times as many parameters, is significantly better. Both OmniX models reach this level with a tenth of the pre-training jets used for OmniLearned. The comparison is not a controlled one, since the models differ in architecture, size, and pre-training data at the same time.

A model that generates jets and events with high fidelity is therefore also a competitive tagger. As in the previous sections, the Events+Jets and the Jets-only model perform equally well. Their $R_{30}$ values agree within the spread of the fine-tunings, and the Events+Jets model has the slightly higher $R_{50}$. Additional training on Collide-2V events does not cost any top-tagging performance.

\section{Conclusions}
\label{sec:conclusions}
We have introduced OmniX, a foundation model that operates on individual
particles and carries out both generative and classification tasks at the
level of single jets and of complete events within a single network. To our knowledge, this is the first time an event-level foundation model operates on three hierarchies at the same time: event level, jet level, and particle level. Two
ingredients make this possible. First, every jet or event is written as one
sequence: conditioning tokens, followed by a coarse plan that holds the
particle multiplicity and summary quantities of up to twenty jets, followed
by the individual particles. This means that the model is encoding both high-level and low-level features of an intrinsically hierarchical event, and is able to attend to everything. Second, the split-and-delay mechanism of
SPADE~\cite{Birk:2026udp} is applied to both the plan and the particles, which keeps
the vocabulary small while still allowing high-resolution features. An added benefit of binning is that it requires no trained tokenizer model. This makes the model flexible in terms of features and datasets. 

OmniX was pre-trained jointly on Collide-2V events with up to twenty
small-radius jets and on single large-radius jets from JetClass, using
next-token prediction and classification as simultaneous objectives. To
isolate the effect of the joint events and jets training, we compared it with versions trained only on jets and only on events. 

In generation, the joint model matches the specialized ones. For full
events, the Events+Jets and the Events-only model agree with the test data to within
a few percent in the bulk of most distributions, and share the same
remaining deviations. For single jets, the results of the Events+Jets and the Jets-only model are nearly indistinguishable. Learning jets and events together therefore
comes at no cost in fidelity.

Classification gives a different picture. When tagging the flavor of all
jets in an event simultaneously, access to the full event improves the separation of
\textit{b} and \textit{c} jets by roughly an order of magnitude compared with a
tagger that sees one jet at a time,
as the model can exploit correlations between the jets of an event. Beyond
this, the Events+Jets model outperforms the Events-only model, although it has seen each
Collide-2V event fewer times. We attribute this gain to the JetClass jets
in the pre-training data, and note that it is based on a single training
run per model. On single-jet tasks, the Events+Jets and the Jets-only model perform
equally: their JetClass accuracy and rejections agree within uncertainties,
and after fine-tuning on the top-tagging dataset both are competitive
with pre-trained models of similar size.
Without any top-tagging labels (zero-shot),
both already exceed the same network trained from scratch on the full top-tagging dataset.

Taken together, a consistent picture emerges. OmniX handles both
representations in a single model, benefits from
global correlations where they are available, and shows transfer from jets to events in flavor tagging. We observe no transfer in
the opposite direction, and none in generation. Whether this changes with
substantially more and more diverse pre-training data, and whether the
flavor-tagging gain is stable across training runs, are natural next
questions.

Event-level models also raise new questions. The non-locality that improves
performance introduces correlations between the objects of an event, which
may make calibration and uncertainty quantification more challenging than
for single-jet taggers. We leave the development of protocols for this
problem to future work.

\section*{Code Availability}
The code is available at \url{https://github.com/uhh-pd-ml/OmniX}.

\section*{Acknowledgments}

The authors thank Thea Aarrestad for support with the Collide-2V dataset. We are also grateful to Louis Moureaux for clarifying discussions during the preparation of this manuscript, and Carsten Burgard for assisting us in data preparation.
The authors acknowledge funding via the SciFM consortium (05D25GU4) funded by the German Federal Ministry of Research, Technology, and Space (BMFTR) in the ErUM-Data action plan, as well as support from the Deutsche Forschungsgemeinschaft (DFG, German Research Foundation) under the German Excellence Initiative -- EXC 2121 Quantum Universe -- 390833306. This work has used the HPC-cluster Hummel-2 at the University of Hamburg. Hummel-2 was funded by Deutsche Forschungsgemeinschaft (DFG, German Research Foundation) -- 498394658.

\section*{Data Availability}
The JetClass dataset is available at \href{https://zenodo.org/records/6619768}{https://zenodo.org/records/6619768}. The Top tagging benchmark dataset is available at \href{https://zenodo.org/records/2603256}{https://zenodo.org/records/2603256}. The Collide-2V dataset is available upon request from the Collide-2V authors~\cite{Moreno:2025collide2v}.

\section*{Declaration on the use of AI tools}
Claude Code was used for coding assistance, under human supervision. 

\bibliography{references}
\bibliographystyle{apsrev4-2}

\appendix
\crefalias{section}{appendix}\crefalias{subsection}{subappendix}
\section{Collide-2V}
\label{app:c2v-data}
\subsection{Preprocessing and selections}
\label{app:c2v-data-processing}
The physical \pt of a particle is the recorded \pt multiplied by the PUPPI weight. Particles with PUPPI weight zero are discarded, as are pileup electrons and muons carrying the \texttt{IsRecoPU} Delphes flag. In the Collide-2V dataset, the $\eta$ and $\phi$ values are stored using \texttt{float16}. With our chosen binning (see \cref{sec:encoding,tab:binning}), this would be too coarse. We therefore dequantize these variables by drawing a value within the \texttt{float16} rounding interval. For charged hadrons, we rebuild the true pseudorapidity $\eta_t$ from the stored $(\pt, \eta, m)$, which agrees better with the generator-level pseudorapidity\footnote{The origin of the mismatch is traced to what variables the PUPPI container passes to FastJet~\cite{Cacciari:2011ma} in Delphes, and is not specific to the Collide-2V dataset.}. The stored pseudorapidity $\eta$ for neutral particles, electrons and muons is used as-is. All jet kinematic quantities are recomputed from the jet constituents, instead of using the stored Delphes values. The jet membership and flavor label from Delphes are kept. The tracker acceptance of the Phase-2 card reaches $|\eta|=4$. We remove the few particles that fall outside of this region.

\begin{figure*}[t]
  \centering
  \includegraphics[width=\linewidth]{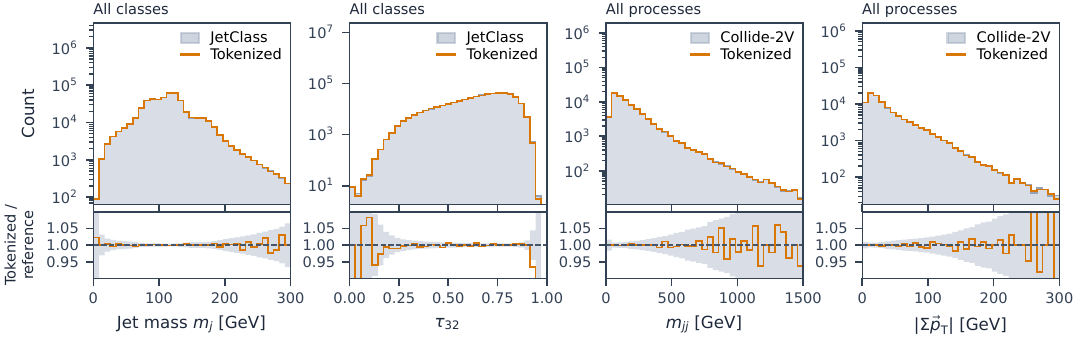}
  \caption{Distributions of test objects before (test data) and after (tokenized) encoding and decoding, for the
  mass and $\tau_{32}$ of \jetclass jets (left), and for the invariant mass $m_{jj}$ of the two leading \akfour
  jets and the magnitude $|\Sigma\vec{p}_\mathrm{T}|$ of the vector sum of the particle transverse momenta in
  Collide-2V events (right). Both curves show the same jets and events, so any difference between them can be attributed to the round trip. The lower panels show the ratio to the test data, with its
  statistical uncertainty as the gray band.}
  \label{fig:codec}
  \vspace{4ex}
  \makeatletter\def\@captype{table}\makeatother
  \caption{The metrics of \cref{tab:metrics-test-jets,tab:metrics-test-events-tt,tab:metrics-test-events-qcd} for the round trip of real jets and
  events through the tokenization, on the samples of \cref{app:discretization}.}
  \label{tab:metrics-test-codec}
  \sisetup{group-digits=none, reset-text-series=false, text-series-to-math=true, propagate-math-font=true}
\begin{tabular*}{\textwidth}{@{\extracolsep{\fill}}l S[table-format=1.3]@{\extracolsep{0pt}${}\pm{}$}S[table-format=1.3]@{\extracolsep{\fill}} S[table-format=-1.2]@{\extracolsep{0pt}${}\pm{}$}S[table-format=1.2]@{\extracolsep{\fill}} S[table-format=1.3] S[table-format=1.2]@{\extracolsep{0pt}${}\pm{}$}S[table-format=1.2]@{\extracolsep{\fill}} S[table-format=2.2]@{\extracolsep{0pt}${}\pm{}$}S[table-format=1.2]@{\extracolsep{\fill}} S[table-format=1.2]@{\extracolsep{0pt}${}\pm{}$}S[table-format=1.2]@{\extracolsep{\fill}} S[table-format=1.2]@{\extracolsep{0pt}${}\pm{}$}S[table-format=1.2]@{\extracolsep{\fill}}}
\toprule
 & \multicolumn{5}{c}{Global} & \multicolumn{8}{c}{Wasserstein distance $W_1$} \\
\cmidrule{2-6}\cmidrule{7-14}
\jetclass jets & \multicolumn{2}{c}{FPD} & \multicolumn{2}{c}{KPD} & \multicolumn{1}{c}{AUC} & \multicolumn{2}{c}{\pt} & \multicolumn{2}{c}{$m_j$} & \multicolumn{2}{c}{$\tau_{21}$} & \multicolumn{2}{c}{$\tau_{32}$} \\
 & \multicolumn{2}{c}{$\times 10^{3}$} & \multicolumn{2}{c}{$\times 10^{5}$} & \multicolumn{1}{c}{} & \multicolumn{2}{c}{GeV} & \multicolumn{2}{c}{GeV} & \multicolumn{2}{c}{$\times 10^{3}$} & \multicolumn{2}{c}{$\times 10^{3}$} \\
\midrule
Truth & 0.070 & 0.022 & -0.1 & 0.4 & 0.501 & 0.25 & 0.12 & 0.13 & 0.04 & 0.39 & 0.24 & 0.52 & 0.15 \\
Tokenized & 0.069 & 0.035 & -0.19 & 0.29 & 0.501 & 0.25 & 0.11 & 0.12 & 0.04 & 0.35 & 0.21 & 0.63 & 0.16 \\
\bottomrule
\noalign{\vskip 8pt}
\toprule
 & \multicolumn{4}{c}{} & \multicolumn{1}{c}{Global} & \multicolumn{8}{c}{Wasserstein distance $W_1$} \\
\cmidrule{6-6}\cmidrule{7-14}
Collide-2V events & \multicolumn{2}{c}{} & \multicolumn{2}{c}{} & \multicolumn{1}{c}{AUC} & \multicolumn{2}{c}{$H_\mathrm{T}$} & \multicolumn{2}{c}{$N_\mathrm{jets}$} & \multicolumn{2}{c}{$p_\mathrm{T}^\mathrm{lead}$} & \multicolumn{2}{c}{$m_{jj}$} \\
 & \multicolumn{2}{c}{} & \multicolumn{2}{c}{} & \multicolumn{1}{c}{} & \multicolumn{2}{c}{GeV} & \multicolumn{2}{c}{$\times 10^{3}$} & \multicolumn{2}{c}{GeV} & \multicolumn{2}{c}{GeV} \\
\midrule
Truth & \multicolumn{2}{c}{} & \multicolumn{2}{c}{} & 0.501 & 1.0 & 0.6 & 11 & 9 & 0.35 & 0.21 & 0.91 & 0.34 \\
Tokenized & \multicolumn{2}{c}{} & \multicolumn{2}{c}{} & 0.502 & 1.0 & 0.6 & 11 & 9 & 0.35 & 0.21 & 0.9 & 0.4 \\
\bottomrule
\end{tabular*}

\end{figure*}

\subsection{Smearing parameters for Collide-2V}
\label{app:c2v-data-smearing}
Smearing is added to the fully unsmeared Collide-2V dataset using values extracted from Fig.~12.12 in Ref.~\cite{CERN-LHCC-2017-009}. The purpose of this added smearing is to give a phenomenologically reasonable starting point that lets us demonstrate the use of the model for flavor tagging. The plots in Ref.~\cite{CERN-LHCC-2017-009} give the resolution in $d_0$ and $d_z$ for $|\eta|<4$ for three \pt values: 1, 10 and 100\,GeV. To cover the \pt range in the Collide-2V dataset better, we perform a fit over the given points. Since we found a joint fit for $|\eta|<2.5$ and $|\eta|>2.5$ to be suboptimal, we perform the fit in \pt separately in these two regions. The points from the three \pt values at $+\eta$ and the three at $-\eta$ are pooled into bins of 0.1 in $|\eta|$. The fitting formula we use for the fit in \pt is
\begin{equation}
    \sigma(p_\mathrm{T}; c) = \left(a_c^{k_r} + \left(\frac{b_c}{p_\mathrm{T}}\right)^{k_r}\right)^{1/{k_r}},
\label{eq:smearing_as_fn_of_pt}
\end{equation}
where $c$ is the bin center in $|\eta|$; $a_c$ and $b_c$ are parameters to be fitted in this bin, and $k_r$ is a parameter to be fitted over the corresponding region $|\eta|\gtrless 2.5$. For each bin $c$ we calculate
\begin{equation}
    \chi^2_c(a_c, b_c; k_r) = \sum_{i\in c}\left(\frac{\log_{10} (\sigma_i/\sigma(p_{\mathrm{T},i};c))}{u_i}\right)^2,
\end{equation}
where $i$ runs over all six points in bin $c$, and $u_i$ is half the error bar on the log scale plots in Ref.~\cite{CERN-LHCC-2017-009} (or half a marker height where the bar is hidden). We then obtain $k_r$ by minimizing over $k$ with the help of the \texttt{minimize\_scalar} function from \texttt{scipy.optimize}~\cite{2020SciPy-NMeth}:
\begin{equation}
    k_r = \arg\min_{k} \sum_{c \in r} \min_{a_c,\, b_c} \chi^2_c(a_c, b_c; k).
\end{equation}
This gives $k_{d_0,|\eta|<2.5} = 1.13$, $k_{d_0,|\eta|>2.5} = 1.05$ and $k_{d_z,|\eta|<2.5} = 1.30$. \Cref{eq:smearing_as_fn_of_pt} is then used to extrapolate below 1\,GeV and above 100\,GeV. For $|\eta|>2.5$ in $d_z$, this fit does not work due to the slope for high \pt changing abruptly around $|\eta|\approx 3.2$. Instead, we approximate this region using a piecewise linear interpolation on the log-log plot. For $p_\mathrm{T}<1$\,GeV, we then continue the resolution as $1/p_\mathrm{T}$. For $p_\mathrm{T}>100$\,GeV, the resolution is kept constant.

\section{Discretization}
\label{app:discretization}

OmniX predicts each continuous feature only as a bin of its codebook (\cref{tab:binning}). To check whether
this limits the fidelity of the generated samples, we encode test data into tokens and decode them exactly as
in generation: continuous values are drawn at random within their bins, and the constituents of each jet are
placed around the jet axis decoded from the plan. This round trip produces what a model that had learned the token
distribution perfectly would generate. Any difference from the input is due to the binning alone. The three
models use the same binning, so the result holds for all of them.

From each dataset we draw two disjoint test samples, of 500\,k \jetclass jets or 100\,k
Collide-2V events each, the latter with the five processes in their dataset proportions. In
\cref{tab:metrics-test-codec}, the
Truth row compares the two unchanged samples, and the Tokenized row compares the second sample after the round trip with the first. The
two rows agree within their uncertainties in every metric, and the distributions in \cref{fig:codec} agree within
the statistical uncertainty of the data. Any deviation of a generated sample beyond the Truth row therefore comes
from the model, not from the discretization.

\begin{figure*}[t]
  \centering
  \includegraphics[width=\linewidth]{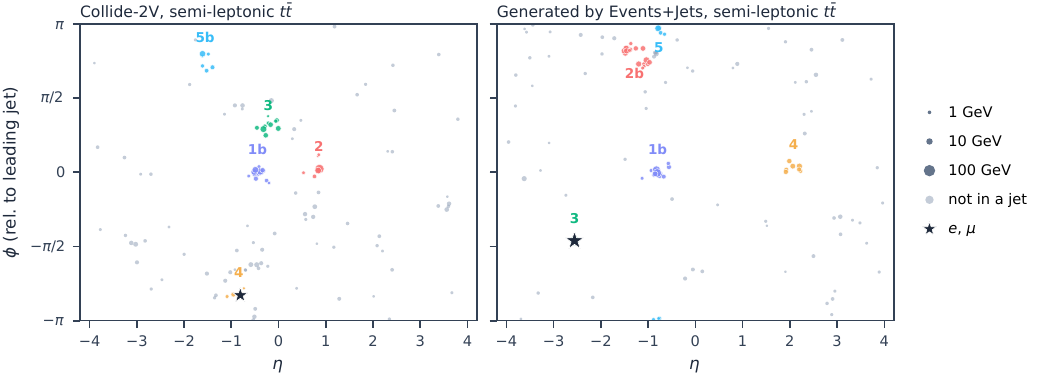}
  \caption{A semi-leptonic \ttbar test event (left) and an event generated by OmniX with the process token
  set to semi-leptonic \ttbar (right). In both examples, the \akfour jets have been clustered with FastJet from the full set of particles in the event. 
  Jets are numbered by decreasing \pt. Gray particles belong to no jet. Jets marked \textit{b} pass the track counting rule described in the text. Stars mark electrons and muons with $\pt>\qty{10}{GeV}$ and the marker area scales
  with $\sqrt{\pt}$. The events were selected by a fixed rule: among the
  events with exactly two jets marked \textit{b}, we take those with the most common jet multiplicity and show the one
  closest to their median $H_\mathrm{T}$.}
  \label{fig:event-display}
\end{figure*}

\section{Additional fidelity comparisons}
\label{app:fidelity}

This appendix shows an example event, examines the jets that the models assign themselves, and compares the generated samples with the test data in more observables than \cref{sec:res-events,sec:res-jets}, together with distance metrics.

\subsection{Example event}
\label{app:fidelity-display}
\Cref{fig:event-display} shows a semi-leptonic \ttbar event generated by OmniX next to an event of the same process from the test set. The generated event looks qualitatively similar to the one from the test set, with several \akfour jets, two of them marked \textit{b}, a lepton, and soft particles over the full range $|\eta|<4$. In both events the jets are clustered from the particles with FastJet~\cite{Cacciari:2011ma} (\antikt~\cite{Cacciari:2008gp}, $R=0.4$, $\pt>\qty{15}{GeV}$, $|\eta|<4$). A jet is marked \textit{b} if it contains at least three charged particles with $|d_0|>3\sigma_{d_0}$, where $\sigma_{d_0}$ is the smearing resolution of \cref{app:c2v-data-smearing}. This simple rule follows the high-purity track counting \textit{b} tag of CMS~\cite{CMS:2012feb}, which uses the impact parameter significance of the third-ranked track in a jet.

\subsection{Reclustered and model-assigned jets}
\label{app:fidelity-modeljets}
\begin{figure*}[t]
  \centering
  \includegraphics[width=\linewidth]{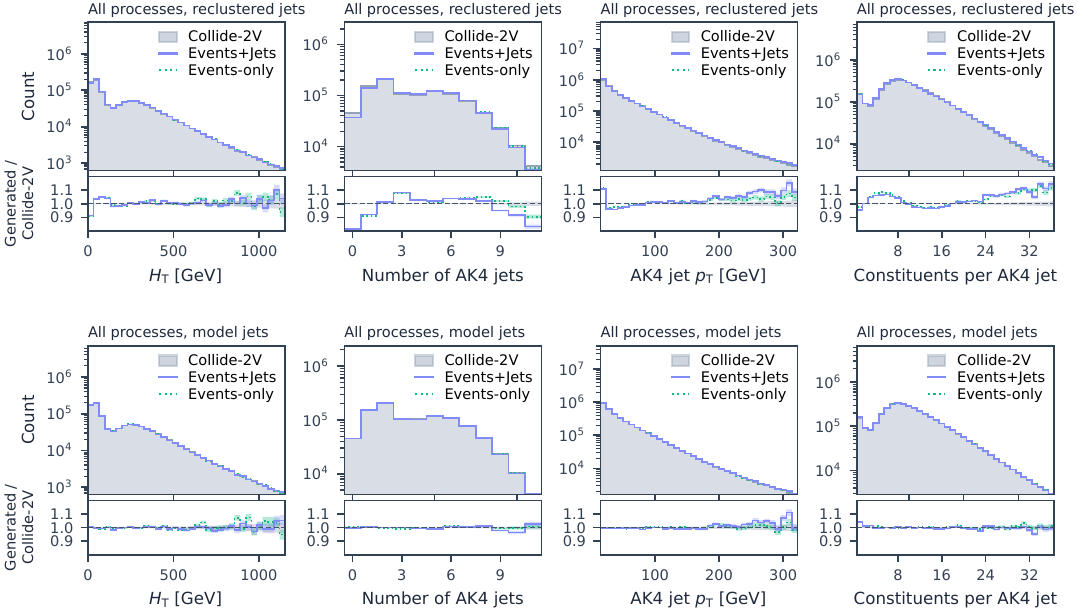}
  \caption{Event and jet observables of the generated Collide-2V events of \cref{fig:events-panels} for two jet
  definitions. In the top row, the \akfour jets are reclustered from all particles with the \antikt algorithm ($R=0.4$, $\pt>\qty{15}{GeV}$, $|\eta|<4$), as in the main text. In the
  bottom row, a jet consists of all particles that carry the same jet index for the
  generated events as assigned by the model and for the test events as given in the dataset. The lower panels
  show the ratio to the Collide-2V test events, with their statistical uncertainty as a gray band.}
  \label{fig:app-c2v-modeljets}
\end{figure*}

The largest deviation in \cref{fig:events-panels} appears in the number of jets. \Cref{fig:app-c2v-modeljets} shows that it arises when the generated particles are reclustered, and that it is absent in the jets that the models generate themselves. As described in \cref{sec:encoding}, every particle carries a jet index, so that a generated event already contains its jets and no clustering is needed to obtain them. \Cref{fig:app-c2v-modeljets} compares the two jet definitions. In the top row, the \akfour jets are reclustered from all particles with the \antikt algorithm ($R=0.4$, $\pt>\qty{15}{GeV}$, $|\eta|<4$), as in the main text. In the bottom row, a jet consists of all particles that carry the same jet index for the generated events as assigned by the model.

With their own jets, both models agree with the test events in all four observables. The models thus generate accurate jets through their plan and the jet index. The deviations in the top row show that these jets do not always match the \akfour definition exactly. Some of them have a \pt below the threshold of \qty{15}{GeV} or contain particles further than $\Delta R=0.4$ from the jet axis, so that the \antikt algorithm groups the generated particles into slightly different jets.

\subsection{Further observables and metrics}
\label{app:fidelity-metrics}
\begin{figure*}[p]
  \centering
  \includegraphics[width=\linewidth,height=0.85\textheight,keepaspectratio]{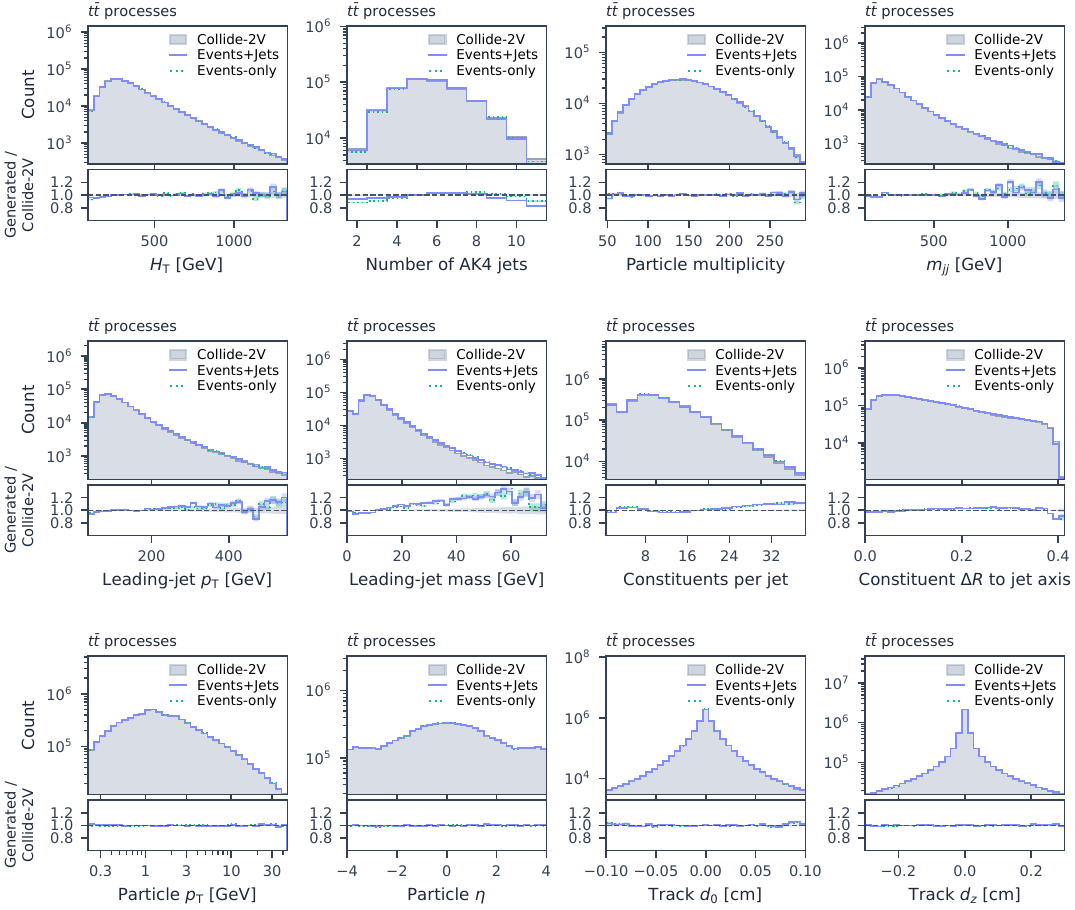}
  \caption{Event, jet, and particle observables of \ttbar events generated by the Events+Jets and the Events-only model
  with the process token set to one of the three \ttbar processes (500\,k events per model, a third per process),
  compared to as many \ttbar test events in the same ratio. $H_\mathrm{T}$ is the scalar sum of the \pt of the
  \akfour jets, and $m_{jj}$ is the invariant mass of the two leading \akfour jets. The constituents per jet and
  their distance $\Delta R$ to the jet axis are shown for all \akfour jets, and $d_0$ and $d_z$ for charged
  particles. The jet-constituent and particle panels show a tenth of the events. The lower panels show the ratio
  to the test events, with the statistical uncertainty of the test events as a gray band.}
  \label{fig:app-c2v-tt}
  \vspace{4ex}
  \makeatletter\def\@captype{table}\makeatother
  \caption{Fidelity of the generated \ttbar events of \cref{fig:app-c2v-tt}, compared with the same test events.
  $W_1$ of the leading-jet \pt and of $m_{jj}$ use the events with at least one and two jets, respectively.}
  \label{tab:metrics-test-events-tt}
  \sisetup{group-digits=none, reset-text-series=false, text-series-to-math=true, propagate-math-font=true}
\begin{tabular*}{\textwidth}{@{\extracolsep{\fill}}l S[table-format=1.3] S[table-format=1.2]@{\extracolsep{0pt}${}\pm{}$}S[table-format=1.2]@{\extracolsep{\fill}} S[table-format=2.1]@{\extracolsep{0pt}${}\pm{}$}S[table-format=1.1]@{\extracolsep{\fill}} S[table-format=1.2]@{\extracolsep{0pt}${}\pm{}$}S[table-format=1.2]@{\extracolsep{\fill}} S[table-format=1.2]@{\extracolsep{0pt}${}\pm{}$}S[table-format=1.2]@{\extracolsep{\fill}} S[table-format=1.2]@{\extracolsep{0pt}${}\pm{}$}S[table-format=1.2]@{\extracolsep{\fill}}}
\toprule
 & \multicolumn{1}{c}{Global} & \multicolumn{10}{c}{Wasserstein distance $W_1$} \\
\cmidrule{2-2}\cmidrule{3-12}
 & \multicolumn{1}{c}{AUC} & \multicolumn{2}{c}{$H_\mathrm{T}$} & \multicolumn{2}{c}{$N_\mathrm{jets}$} & \multicolumn{2}{c}{$p_\mathrm{T}^\mathrm{lead}$} & \multicolumn{2}{c}{$m_{jj}$} & \multicolumn{2}{c}{$N_\mathrm{particles}$} \\
 & \multicolumn{1}{c}{} & \multicolumn{2}{c}{GeV} & \multicolumn{2}{c}{$\times 10^{3}$} & \multicolumn{2}{c}{GeV} & \multicolumn{2}{c}{GeV} & \multicolumn{2}{c}{} \\
\midrule
Truth & 0.502 & 0.47 & 0.26 & 6.6 & 2.0 & 0.18 & 0.06 & 0.58 & 0.14 & 0.17 & 0.06 \\
Events+Jets & 0.571 & 5.2 & 0.5 & 44.0 & 2.3 & 4.93 & 0.24 & 4.7 & 0.5 & 0.15 & 0.04 \\
Events-only & 0.558 & 5.9 & 0.4 & 58.8 & 2.2 & 4.59 & 0.23 & 4.2 & 0.4 & 0.18 & 0.06 \\
\bottomrule
\end{tabular*}

\end{figure*}

\begin{figure*}[p]
  \centering
  \includegraphics[width=\linewidth,height=0.85\textheight,keepaspectratio]{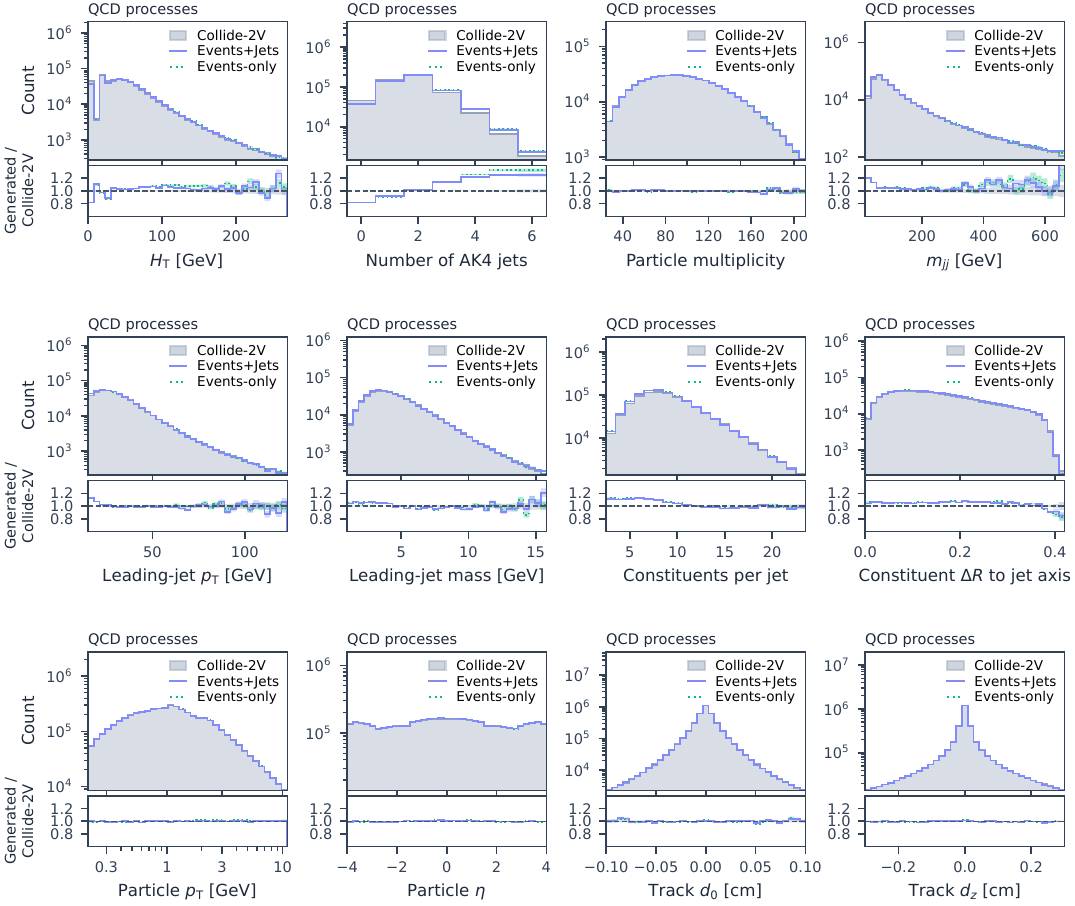}
  \caption{As in \cref{fig:app-c2v-tt}, for QCD events generated with the process token set to inclusive QCD or
  QCD~$\to b\bar{b}$ (500\,k events per model, in the ratio of the training data).}
  \label{fig:app-c2v-qcd}
  \vspace{4ex}
  \makeatletter\def\@captype{table}\makeatother
  \caption{As in \cref{tab:metrics-test-events-tt}, for the QCD events of \cref{fig:app-c2v-qcd}.}
  \label{tab:metrics-test-events-qcd}
  \sisetup{group-digits=none, reset-text-series=false, text-series-to-math=true, propagate-math-font=true}
\begin{tabular*}{\textwidth}{@{\extracolsep{\fill}}l S[table-format=1.3] S[table-format=1.2]@{\extracolsep{0pt}${}\pm{}$}S[table-format=1.2]@{\extracolsep{\fill}} S[table-format=3.1]@{\extracolsep{0pt}${}\pm{}$}S[table-format=1.1]@{\extracolsep{\fill}} S[table-format=1.3]@{\extracolsep{0pt}${}\pm{}$}S[table-format=1.3]@{\extracolsep{\fill}} S[table-format=1.2]@{\extracolsep{0pt}${}\pm{}$}S[table-format=1.2]@{\extracolsep{\fill}} S[table-format=1.3]@{\extracolsep{0pt}${}\pm{}$}S[table-format=1.3]@{\extracolsep{\fill}}}
\toprule
 & \multicolumn{1}{c}{Global} & \multicolumn{10}{c}{Wasserstein distance $W_1$} \\
\cmidrule{2-2}\cmidrule{3-12}
 & \multicolumn{1}{c}{AUC} & \multicolumn{2}{c}{$H_\mathrm{T}$} & \multicolumn{2}{c}{$N_\mathrm{jets}$} & \multicolumn{2}{c}{$p_\mathrm{T}^\mathrm{lead}$} & \multicolumn{2}{c}{$m_{jj}$} & \multicolumn{2}{c}{$N_\mathrm{particles}$} \\
 & \multicolumn{1}{c}{} & \multicolumn{2}{c}{GeV} & \multicolumn{2}{c}{$\times 10^{3}$} & \multicolumn{2}{c}{GeV} & \multicolumn{2}{c}{GeV} & \multicolumn{2}{c}{} \\
\midrule
Truth & 0.501 & 0.10 & 0.04 & 1.7 & 0.8 & 0.059 & 0.015 & 0.40 & 0.20 & 0.090 & 0.019 \\
Events+Jets & 0.572 & 3.28 & 0.16 & 113.7 & 1.6 & 1.66 & 0.04 & 2.72 & 0.27 & 0.19 & 0.05 \\
Events-only & 0.563 & 3.61 & 0.13 & 126.8 & 2.0 & 1.65 & 0.06 & 2.37 & 0.17 & 0.108 & 0.031 \\
\bottomrule
\end{tabular*}

\end{figure*}

\begin{figure*}[p]
  \centering
  \includegraphics[width=\linewidth,height=0.85\textheight,keepaspectratio]{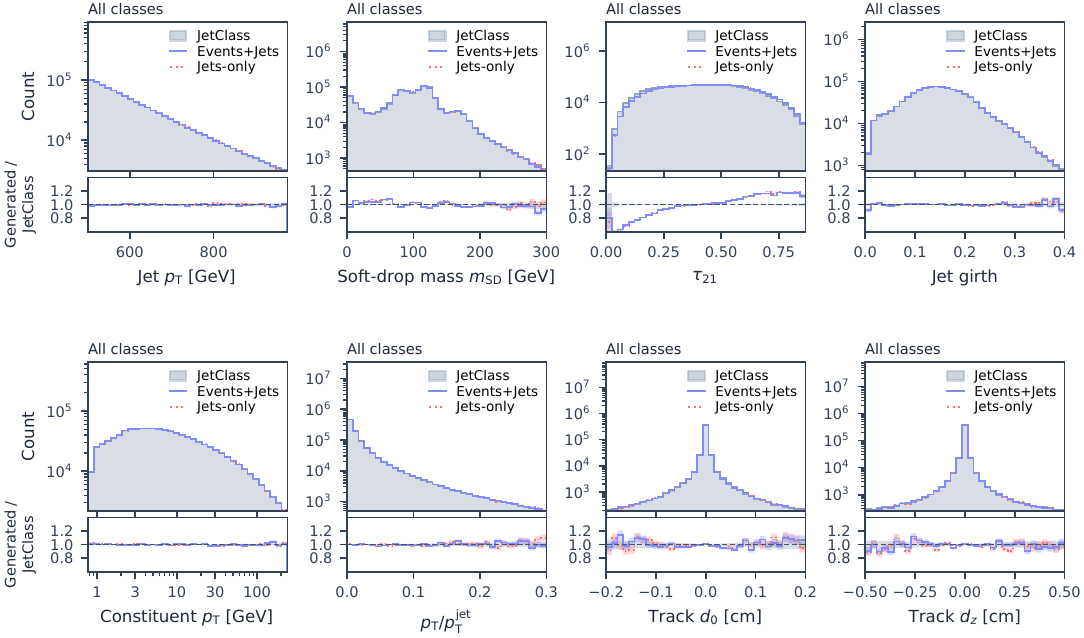}
  \caption{Jet-level and constituent-level observables of \jetclass jets generated by the Events+Jets and the Jets-only model.}
  \label{fig:app-jets-jet}
  \vspace{4ex}
  \makeatletter\def\@captype{table}\makeatother
  \caption{Fidelity of generated \jetclass jets. Each model generates 1\,M jets, 100\,k with the process
  token set to each of the ten classes. They are compared with an equal number of test jets of all ten classes.}
  \label{tab:metrics-test-jets}
  \sisetup{group-digits=none, reset-text-series=false, text-series-to-math=true, propagate-math-font=true}
\begin{tabular*}{\textwidth}{@{\extracolsep{\fill}}l S[table-format=1.3]@{\extracolsep{0pt}${}\pm{}$}S[table-format=1.3]@{\extracolsep{\fill}} S[table-format=-1.2]@{\extracolsep{0pt}${}\pm{}$}S[table-format=1.2]@{\extracolsep{\fill}} S[table-format=1.3] S[table-format=1.2]@{\extracolsep{0pt}${}\pm{}$}S[table-format=1.2]@{\extracolsep{\fill}} S[table-format=1.2]@{\extracolsep{0pt}${}\pm{}$}S[table-format=1.2]@{\extracolsep{\fill}} S[table-format=1.3]@{\extracolsep{0pt}${}\pm{}$}S[table-format=1.3]@{\extracolsep{\fill}} S[table-format=2.2]@{\extracolsep{0pt}${}\pm{}$}S[table-format=1.2]@{\extracolsep{\fill}} S[table-format=2.2]@{\extracolsep{0pt}${}\pm{}$}S[table-format=1.2]@{\extracolsep{\fill}}}
\toprule
 & \multicolumn{5}{c}{Global} & \multicolumn{10}{c}{Wasserstein distance $W_1$} \\
\cmidrule{2-6}\cmidrule{7-16}
 & \multicolumn{2}{c}{FPD} & \multicolumn{2}{c}{KPD} & \multicolumn{1}{c}{AUC} & \multicolumn{2}{c}{\pt} & \multicolumn{2}{c}{$m_j$} & \multicolumn{2}{c}{$N_\mathrm{const}$} & \multicolumn{2}{c}{$\tau_{21}$} & \multicolumn{2}{c}{$\tau_{32}$} \\
 & \multicolumn{2}{c}{$\times 10^{3}$} & \multicolumn{2}{c}{$\times 10^{5}$} & \multicolumn{1}{c}{} & \multicolumn{2}{c}{GeV} & \multicolumn{2}{c}{GeV} & \multicolumn{2}{c}{} & \multicolumn{2}{c}{$\times 10^{3}$} & \multicolumn{2}{c}{$\times 10^{3}$} \\
\midrule
Truth & 0.051 & 0.026 & -0.21 & 0.12 & 0.500 & 0.10 & 0.06 & 0.14 & 0.06 & 0.021 & 0.005 & 0.19 & 0.11 & 0.20 & 0.14 \\
Events+Jets & 1.8 & 0.7 & -0.18 & 0.22 & 0.527 & 1.02 & 0.12 & 0.51 & 0.05 & 0.025 & 0.008 & 17.28 & 0.29 & 10.55 & 0.13 \\
Jets-only & 2.0 & 0.6 & -0.06 & 0.24 & 0.527 & 1.33 & 0.13 & 0.80 & 0.11 & 0.019 & 0.011 & 17.66 & 0.28 & 11.16 & 0.25 \\
\bottomrule
\end{tabular*}

\end{figure*}

\Cref{fig:app-c2v-tt,fig:app-c2v-qcd} and \cref{tab:metrics-test-events-tt,tab:metrics-test-events-qcd} split the generated Collide-2V events of \cref{fig:events-panels} by the process token they were generated with into the three \ttbar and the two QCD processes, and compare each group with as many test events of the same processes. \Cref{fig:app-jets-jet} and \cref{tab:metrics-test-jets} show \jetclass jets, generated with the process token set to each of the ten classes in equal parts. In the generated and in the test events, \akfour jets are clustered from all particles with the \antikt algorithm~\cite{Cacciari:2008gp,Cacciari:2011ma} with $R=0.4$ and kept if they satisfy $\pt>\qty{15}{GeV}$ and $|\eta|<4$.

All tables quote the same metrics. The Wasserstein-1 distance $W_1$ compares the generated and the test distribution of a single observable and is given in the unit of that observable. Its uncertainty is the standard deviation over ten bootstrap resamplings of both samples. The other metrics compare the samples in several observables at once. The AUC is the area under the ROC curve of a gradient-boosted decision tree that is trained to separate generated from test objects and evaluated on a held-out half of the two samples. An AUC of 0.5 means that the classifier cannot tell them apart. For jets, the classifier uses eight observables: the jet \pt, the mass, the number of constituents, the girth $\sum_i p_{\mathrm{T},i}\,\Delta R_i / p_\mathrm{T}^\mathrm{jet}$, the momentum dispersion $p_\mathrm{T}^D = \sqrt{\sum_i p_{\mathrm{T},i}^2}/\sum_i p_{\mathrm{T},i}$, the number of charged constituents, and the \pt fractions carried by the charged constituents and by the leading constituent. For jets we also calculate the Fr\'echet physics distance (FPD) and the kernel physics distance (KPD)~\cite{Kansal:2022spb}, computed with the \textsc{JetNet} library~\cite{Kansal:2023iqy} on energy flow polynomials~\cite{Komiske:2017aww}. Both are established for jets only, so we do not quote them for events. A small fraction of generated jets, 0.3\% for both models, has at least one of the eight observables outside the range covered by the test jets. These jets are removed before FPD, KPD, and the AUC are computed, because a few extreme jets would otherwise dominate the FPD. The $W_1$ distances use all jets. In every table, the Truth row compares a second, independent sample of test objects with the first and gives the value that each metric takes for two samples from the same distribution. For events, the gradient-boosted decision tree is trained on the 22 features listed in \cref{tab:event-observables}.

\begin{table}[t]
  \caption{The 22 event observables used by the classifier for Collide-2V events. Jet observables refer to the \akfour jets ordered in \pt, and quantities of absent jets are set to zero.}
  \label{tab:event-observables}
  \begin{tabular}{ll}
    \toprule
    Level & Observables \\
    \midrule
    Particles (10) & multiplicity of all particles \\
     & multiplicity of charged particles, photons, \\
     & electrons, and muons \\
     & multiplicity of tracks with $|d_0|>3\sigma_{d_0}$ \\
     & scalar \pt sum and $|\Sigma\vec{p}_\mathrm{T}|$ \\
     & leading-particle \pt \\
     & fraction of the scalar \pt sum at $|\eta|>2.5$ \\
    \midrule
    Jets (9) & number of jets \\
     & \pt and mass of the two leading jets \\
     & $|y|$ of the leading jet, $\Delta\phi_{12}$, $m_{jj}$ \\
     & $A_{12}=(p_\mathrm{T,1}-p_\mathrm{T,2})/(p_\mathrm{T,1}+p_\mathrm{T,2})$ \\
    \midrule
    Out-of-jet (3) & multiplicity, \pt sum, and \pt fraction \\
    \bottomrule
  \end{tabular}
\end{table}

\clearpage
\section{Hyperparameters}
\label{app:hyperparameters}
\Cref{tab:hyper-shared} lists the hyperparameters shared by all three models. The settings that differ between the three models, including the number of training steps, are given in \cref{tab:models}. No extensive hyperparameter optimization was performed.
\Cref{tab:hyper-finetune} lists the settings of the fine-tuning on top tagging in \cref{sec:res-finetune}.
\onecolumngrid
\begin{center}
\begin{minipage}{\textwidth}
  \centering
  \makeatletter\def\@captype{table}\makeatother
  \caption{Hyperparameters shared by all three OmniX models. $S$ is the length of the
  constant learning-rate phase.}
  \label{tab:hyper-shared}
  \begin{tabular}{lll}
    \toprule
    Group & Parameter & Value \\
    \midrule
    \multirow{7}{*}{Sequence}
      & maximum number of particles & \nMaxPartCv\ (Collide-2V), \nMaxPartJC\ (\jetclass) \\
      & maximum sequence length & 704 \\
      & particle order & random, redrawn each time an event is seen \\
      & delay orders & 32 of 420 admissible \\
      & feature mode & full or kinematic, probability 0.5 each \\
      & process token given & probability 0.5 in training, never in validation \\
      & conditioning + plan positions & $4+25$ \\
    \midrule
    \multirow{4}{*}{\shortstack[l]{Conditioning\\codebooks}}
      & dataset & 2 (\jetclass, Collide-2V) \\
      & feature mode & 2 (full, kinematic) \\
      & delay order & 32 \\
      & process & 16 (\texttt{NULL}, 10 \jetclass classes, 5 Collide-2V processes) \\
    \midrule
    \multirow{7}{*}{Backbone}
      & blocks & 8, pre-layer-norm, final layer norm \\
      & width & 768 \\
      & attention & multi-query, 8 heads of dimension 96 \\
      & position encoding & rotary \\
      & feed-forward & SwiGLU, hidden size 3072 \\
      & dropout & 0 \\
      & parameters & \nParBackbone~M \\
    \midrule
    \multirow{4}{*}{Inputs}
      & token embedding & $7\times64$, layer norm, linear $448\to768$ \\
      & continuous input MLP & $3\to128\to512\to768$, GELU~\cite{Hendrycks:2016gelu} \\
      & counter MLP & $7\to64\to768$, SiLU~\cite{Elfwing:2018silu}, zero-initialized output \\
      & pairwise-bias MLP & $8\to64\to64\to64\to8$, SiLU, zero-initialized output \\
    \midrule
    \multirow{5}{*}{Heads}
      & generative & one linear layer per particle feature and per plan quantity \\
      & jet-class (\jetclass) & 2 class-attention blocks (8 heads, SwiGLU 3072) \\
      & & MLP $768\to256\to256\to10$, SiLU (\nParHeadJC~M) \\
      & jet-flavor (Collide-2V) & 3 blocks (self- and cross-attention, SwiGLU) \\
      & & MLP $768\to256\to256\to3$, SiLU (\nParHeadCv~M) \\
    \midrule
    \multirow{9}{*}{Training}
      & optimizer & Ranger (RAdam + Lookahead) \\
      & $\beta_1$, $\beta_2$, $\epsilon$ & 0.95, 0.999, $10^{-5}$ \\
      & Lookahead $\alpha$, $k$ & 0.25, 16 \\
      & weight decay & $5\times10^{-3}$, decoupled, on all parameters \\
      & gradient clipping & 1.0 (norm) \\
      & batch size & 1024 events ($8\times128$ on 8 H100 GPUs) \\
      & precision & bf16 mixed \\
      & loss weights & 1 for every term \\
    \midrule
    \multirow{3}{*}{\shortstack[l]{Learning\\rate}}
      & warm-up & linear to $10^{-3}$ over 4096 steps \\
      & constant phase & $10^{-3}$, at most $S=\num{2457600}$ steps \\
      & anneal & cosine to $5\times10^{-6}$ over $0.2\,S$, rounded up to a multiple of 8192 steps \\
    \midrule
    \multirow{2}{*}{Generation}
      & sampling & temperature 1, key-value cache \\
      & delay order, feature mode & uniform over the codebook; full mode \\
    \bottomrule
  \end{tabular}
\end{minipage}
\end{center}
\twocolumngrid

\begin{table*}[t]
  \caption{Settings of the fine-tuning on top tagging in \cref{sec:res-finetune}. Where four values are given, they apply to
  1\,k, 10\,k, 100\,k, and 1.2\,M labeled training jets.}
  \label{tab:hyper-finetune}
  \begin{tabular}{lll}
    \toprule
    Group & Parameter & Value \\
    \midrule
    \multirow{4}{*}{Input}
      & dataset token & \jetclass \\
      & feature mode & kinematic \\
      & process token & \texttt{NULL} \\
      & continuous inputs & exact values, not dithered \\
    \midrule
    \multirow{2}{*}{Initialization}
      & network & pre-trained Events+Jets or Jets-only model, or random (from scratch) \\
      & top-tagging head & jet-class head, outputs for QCD and $t\to bqq'$ \\
    \midrule
    \multirow{9}{*}{Training}
      & loss & cross-entropy of the classification pass, generative pass not used \\
      & optimizer & AdamW, $\beta_1=0.95$, $\beta_2=0.999$, $\epsilon=10^{-5}$, weight decay 0.01 \\
      & peak learning rate & $10^{-4}$ for the MLP and query of the head, $10^{-5}$ for all other weights \\
      & peak learning rate, from scratch & $10^{-4}$ / $3\times10^{-4}$ / $3\times10^{-4}$ / $10^{-4}$ for all weights \\
      & schedule & linear warm-up over 10\% of the steps, then cosine to 5\% of the peak \\
      & batch size & 100 / 100 / 512 / 512 \\
      & steps & \num{1000} / \num{2000} / \num{4000} / \num{40000} \\
      & passes over the labeled jets & 100 / 20 / 20 / 17 \\
      & gradient clipping, precision, dropout & 1.0 (norm), bf16 mixed, 0 \\
    \midrule
    \multirow{4}{*}{Selection}
      & validation set & about \num{40000} jets \\
      & validation every & 10 / 100 / 200 / 500 steps \\
      & checkpoint & highest validation AUC \\
      & runs & 5 per training set size \\
    \bottomrule
  \end{tabular}
\end{table*}

\end{document}